%% file: main.tex
\documentclass[11pt]{article}

\input{preamble.tex}

\title{Coacervation drives morphological diversity of mRNA encapsulating nanoparticles}
\date{\today}

\usepackage{subcaption, textcomp, textgreek}

\def\d{\textrm{d}}

\graphicspath{{figures/}}
\usepackage[sort&compress, numbers]{natbib}

\begin{document}

\author{Emmit K. Pert, Paul J. Hurst, Robert M. Waymouth, and Grant M. Rotskoff \\
Department of Chemistry, Stanford University, Stanford, CA USA 94305}

\maketitle

\begin{abstract}
    The spatial arrangement of components within an mRNA encapsulating nanoparticle has consequences for its thermal stability, which is a key parameter for therapeutic utility.
    The mesostructure of mRNA nanoparticles formed with cationic polymers have several distinct putative structures: here, we develop a field theoretic simulation model to compute the phase diagram for amphiphilic block copolymers that balance coacervation and hydrophobicity as driving forces for assembly. 
    We predict several distinct morphologies for the mesostructure of these nanoparticles, depending on salt conditions and hydrophobicity. 
    We compare our predictions with cryogenic-electron microscopy images of mRNA encapsulated by charge altering releasable transporters.
    In addition, we provide a GPU-accelerated, open-source codebase for general purpose field theoretic simulations, which we anticipate will be a useful tool for the community. 
\end{abstract}

\section{Introduction}

The widespread deployment of lipid nanoparticle mRNA vaccines~\cite{aldosari_lipid_2021,hou_lipid_2021} has accelerated efforts to diversify polymer-based drug delivery technologies.
A number of physical properties, including high thermal stability and structural robustness~\cite{hou_lipid_2021}, are desirable for therapeutics, but we currently lack the necessary design principles to ensure these features on theoretical grounds~\cite{chan_computational_2021}.
Among the many candidate technologies~\cite{hou_lipid_2021}, amphiphilic polymers that encapsulate mRNA via coacervation have emerged as a potentially fruitful route to building nanoparticles with delivery specificity~\cite{hou_lipid_2021,aldosari_lipid_2021,mckinlay_charge-altering_2017,mckinlay_enhanced_2018, li_chargealtering_2023}. 
mRNA encapsulating nanoparticles are also of interest as simplified models for liquid-liquid phase separation in biological systems \cite{brangwynne_germline_2009}. 
Here, we study the structural properties of mRNA encapsulating polymeric nanoparticles using field theoretic simulations and cryogenic-electron microscopy (cryoEM) to elucidate the physiochemical driving forces that underlie several distinct mesostructures.

While ionizable lipid nanoparticles have been the dominant modality for non-viral gene delivery vectors~\cite{schoenmaker_mrnalipid_2021}, a number of cationic synthetic polymers have emerged as alternatives for mRNA encapsulation and delivery~\cite{liu_mrna_2023, huang_roles_2022, bezbaruah_nanoparticle-based_2022, mckinlay_charge-altering_2017}.
In this work, we focus on charge altering releasable transports (CARTs), a class of copolymers formed from a cationic block and lipid block that exhibits a remarkable self-immolative degradation to neutral small molecule, hypothesized to aid in endosomal escape of mRNA~\cite{mckinlay_charge-altering_2017}, though we study only the thermodynamics of the pre-degradation polymers in this work. 
Distinct formulations of CART polymers have been demonstrated to produce delivery vehicles that allow for targeting of distinct cell types or even preferentially target individual organs~\cite{mckinlay_charge-altering_2017,mckinlay_enhanced_2018,li_organ_2024}.
However, assessing the impact of tuneable parameters, like block length and hydrophobicity of the encapsulating polymers on the self-organization of the resulting nanoparticles remains a largely empirical effort due to the absence of predictive models~\cite{chan_computational_2021}.
Furthermore, the incipient understanding of tissue selectivity has not yet been unambiguously connected to structural features of delivery nanoparticles~\cite{li_organ_2024}. 

The physical mechanism that drives assembly of CARTs features both a hydrodynamic driving force and electrostatic interactions between polyanionic mRNA and the charged block of the encapsulating polymers~\cite{mckinlay_charge-altering_2017} via a process known as coacervation.
Coacervation is an active area of research, both experimentally and theoretically~\cite{rumyantsev_polyelectrolyte_2021}.
Because coacervation is a fluctuation-induced effect, mean-field theories incorrectly predict that no assembly of oppositely charged polymers should occur~\cite{sing_development_2017,delaney_theory_2017}.
Analytical theories that go beyond mean-field, such as the random phase approximation (RPA), make corrections that resolve the deficiency of mean-field theory, and predict correct scaling behavior~\cite{borue_statistical_1990, nakamura_thermodynamics_2011, dobrynin_scaling_1995}.
However, it remains challenging within analytical theory to simultaneously incorporate the effect of hydrophobicity, charge, and sequence and more sophisticated models that represent dynamical fluctuations are needed~\cite{sing_electrostatic_2014,sing_bridging_2023}.

While some progress on coacervation dynamics has been made using atomistic~\cite{sing_development_2017, velichko_molecular_2008} or coarse-grained~\cite{rumyantsev_polyelectrolyte_2021, yu_complex_2021} models, the computational costs of simulating mesostructure is prohibitive for many systems of interest, so we turn to field theoretic simulation~\cite{fredrickson_field-theoretic_2023} of a polymer field theory~\cite{fredrickson_equilibrium_2006}.
Numerical simulation of an appropriate polymer field theory offers an appealing alternative to molecular models for systems that balance coacervation with hydrophobic driving forces, though to our knowledge this specific setting has not been  previously studied. 
Because polymer field theories map the density of each species to a continuous field, they can be employed to predict both the spatial organization of components and also thermodynamic properties derivable from the computationally accessible partition function~\cite{fredrickson_equilibrium_2006}. 
There is a vast literature on numerical simulation of polymer field theories, discussed at length in the recent monograph Ref.~\cite{fredrickson_field-theoretic_2023}, which has led to algorithms that make field theoretic simulation computationally efficient and demonstrations that numerical simulations of polymer field theories yield excellent agreement with experimentally determined morphology in a variety of contexts~\cite{najafi_liquidliquid_2021, lindsay_complex_2021, albanese_improved_2023}.
Despite its suitability for the self-organization of polymers like CART-mRNA nanoparticles, previous work on coacervation has focused primarily on the setting of a ``good'' solvent to isolate electrostatic effects~\cite{rumyantsev_polyelectrolyte_2021}.

Because we focus on the generic physical properties that drive encapsulation in this work, we develop a field theoretic representation of the CART-mRNA nanoparticle system, a highly coarse-grained representation, that enables simulations at the ``mesoscale''. 
We find that this model predicts morphological features also present in experimental cryoEM. 
We observe a transition to a dense coacervate phase, the specific morphology of which depends strongly on both salt concentration and polymer hydrophobicity.
Despite minimal modeling assumptions, the morphologies that we observe are in excellent agreement with experimental cryoEM micrographs. 
The methodology and open-source, GPU-accelerated, simulation package that we present here are both broadly applicable to polymer mixtures in solvent; we anticipate, in particular, that our software package will be useful for the polymer physics community because existing open-source tools for field theoretic simulation do not include fluctuations beyond static self-consistent field theory~\cite{cheong_open-source_2020}.

\begin{figure}
    \centering
    \includegraphics[width=0.5\linewidth]{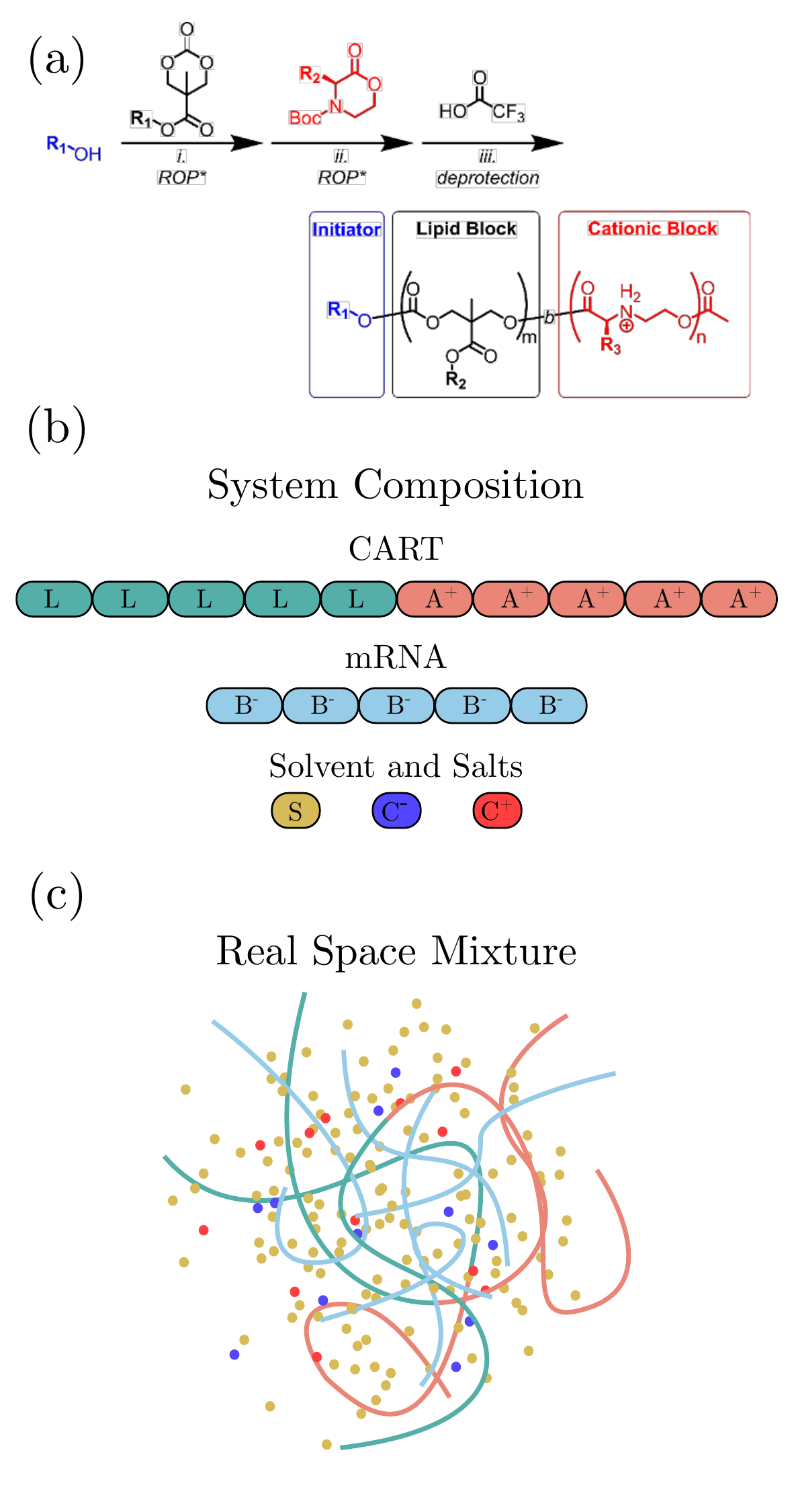}
    \caption{Description of system composition and schematic of mixture in real space. (a) CART preparation and structure containing an initiator (R1) lipophilic carbonate block with lipid side chain (R2), and a hydrophilic cationic block derived from N-hydroxyethyl $\alpha$-amino acids (R3 for amino acid side chain).  (b) Structure of all components (L - lipid block, A - cationic hydrophilic block, B - anionic hydrophobic block) (c) Schematic showing mixture of solvent, salt, and polymers}
    \label{fig:schematic}
\end{figure}

\section{Theoretical Framework}

Throughout, we consider a mixture of charged and neutral polymers in solvent with explicit salt.
There are $M$ total polymer species and each species has $n_m$ polymers.
An individual polymer takes coordinates $\rb \in \Omega \subset \mathbb{\RR}^d$, a domain with volume $V$ and periodic boundary conditions.
In this work $d=2,$ though all equations are general for any $d$.
We first write the explicit polymer density for species $m$
as 
\begin{equation}
    \bar \rho^{(m)} (\rb) = \sum_{n=1}^{n_m} \int_0^{N_m/N} \delta (\rb - \rb_n^{(m)}(s))\ \d s
\end{equation} 
where $N_m$ is the length of a polymer of species $m.$
This configuration dependent density is singular \cite{villet_efficient_2014}, so we convolve with a Gaussian kernel to resolve the ultraviolet divergence,
\begin{equation}
    \Gamma(\rb) = (2 \pi a^2)^{d/2} \exp \left(-\frac{1}{2 a^2} \rb^T \rb \right),
    \label{eq:gauss}
\end{equation}
to obtain
\begin{equation}
     \rho^{(m)} (\rb) \equiv \Gamma * \bar \rho^{(m)} (\rb) = \int_{\Omega} \Gamma(| \rb - \rb'|) \bar \rho^{(m)} (\rb) \ \d \rb.
\end{equation}
Similarly, we write the smoothed charge density for species $m$ as
\begin{equation}
     \rho_{\rm C}^{(m)} (\rb) = \Gamma * \bar \rho_{\rm C}^{(m)} (\rb)
\end{equation}
with 
\begin{equation}
    \bar \rho_{\rm C}^{(m)} (\rb) = \sum_{n=1}^{n_m} \int_0^{N_m / N} s_n(s) \delta (\rb - \rb_n^{(m)}(s))\ \d s.
\end{equation}
where $s:[0,N_m / N] \to \RR$ gives the charge along the polymer.
We denote the vector of all monomer densities $\rhob:\Omega \to \RR^M$ and the total charge density is $\rho_{\rm C}(\rb) \equiv \sum_{i=0}^M Z_i \rho_i (\rb)$, where $Z_i$ is the charge of species $i$.

Assuming local pairwise interactions and Coulomb charge-charge interactions, the Hamiltonian for the system is
\begin{equation}
    \beta U[\rhob, \rhob_{\rm C}] = \int_{\Omega} \rhob(\rb)^T \chi \rhob(\rb) \d \rb \d \rb' + \frac{\ell_{\rm B}}{2} \int_{\Omega} \frac{\rho_{\rm C}(\rb) \rho_{\rm C}(\rb')}{|\rb - \rb'|} \d \rb \d \rb' - \beta U_0.
\end{equation}
Here $\chi\in \RR^{M \times M}$ is the Flory-Huggins (FH) interaction matrix and $\ell_{\rm B}$ is the Bjerrum length; $U_0$ is a constant arising from the polymer self-interaction. 
Following standard arguments~\cite{fredrickson_equilibrium_2006}, we can express the canonical partition function as a field theory with a complex-valued Hamiltonian $\mathcal{H}$, 
\begin{equation}
    \mathcal{Z}(\beta) = \mathcal{Z}_{\rm ideal} \prod_{m=1}^{M+S} \int \mathcal{D} \omega_m \mathcal{D} \varphi e^{-\mathcal{H}[\{ \omega \}, \varphi]}
    \label{eq:parti}
\end{equation}
where 
\begin{equation}
\begin{aligned}
    \mathcal{H}[\{\omega\}, \varphi] &= \sum_{m=1}^M \frac{\gamma_i^2}{2B_i} \int_{\Omega} \omega_m^2(\rb) \d \rb + \frac{1}{2E} \int_{\Omega} | \nabla \varphi(\rb) |^2 \d \rb \\
    & -\sum_{i=1}^{P+S} n_i \log Q_i[\{\omega \}, \varphi].
    \label{eq:ham}
\end{aligned}
\end{equation}
The rescaled Flory-Huggins interaction in the diagonal basis is given by $(2B_i)^{-1}$ and $\gamma_i = i$ if $B_i < 0$ and is unity otherwise.
The parameter $E$ is related to the Bjerrum length and is explicitly $E = 4\pi L^2 \ell_{\rm B} / R_{\rm g}.$
This expression has a straightforward interpretation in the simplest settings: the $\omega_m$ are chemical potential fields associated with the $M$ polymer density fields and the $S$ solvent density fields, and $\varphi$ is the field conjugate to the charge density, which can be interpreted as an electrostatic potential field. We emphasize in our case, as detailed in Appendix~\ref{app:deriv}, the pairwise interactions between density fields are orthonormalized so throughout $\omega_m$ should be interpreted as the chemical potential for a linear combination of polymer and solvent fields. 

The simplest way to approximately solve \eqref{eq:parti} is to compute the mean-field solution self-consistently. 
Unfortunately, this approach is incapable of predicting fluctuation-induced effects, including coacervation.
When the Flory-Huggins interactions between all pairs of species are equal, a straightforward mean-field treatment of polyelectrolytes incorrectly converges to a homogeneous distribution of material, so including fluctuations is crucial~\cite{delaney_theory_2017}.
To accurately incorporate the effect of thermal fluctuations, stochastic sampling techniques such as a Monte Carlo sampling or field theoretic simulation (FTS) must be employed. 
For statistical field theories like the polymer field theory represented with the Hamiltonian~\eqref{eq:ham}, FTS has already been widely used~\cite{delaney_recent_2016} and primarily relies on the complex Langevin algorithm~\cite{fredrickson_equilibrium_2006}.  
We describe our implementation of this algorithm in Appendix~\ref{app:CL}.

The systems we consider are incompressible polymer mixtures in solvent.
To enforce incompressibility, previous work imposed this constraint via $\omega_+$, the field associated with total density~\cite{duchs_multi-species_2014}. 
Specifically, the Hamiltonian removes the $\omega_+^2$ term, allowing $\omega_+$ to exert whatever chemical potential is needed to maintain constant spatial density. 
While this approach imposes the constraint exactly, it leads to additional numerical stiffness in the system.
Instead, we allow $\omega_+$ to fluctuate with a weak density constraint like all other fields in the system, sometimes called a ``soft explicit solvent''.
This name indicates that, while the solvent field is explicitly included, the dynamics allows violations of homogeneous total density. 
An alternative, physically consistent interpretation presents the explicit solvent as permitting attractive interactions, while an implicit co-solvent that has only repulsive interactions fills in any missing density.

\section{Results and Discussion}

\begin{figure}[ht]
    \centering
    \includegraphics[width=0.99\linewidth]{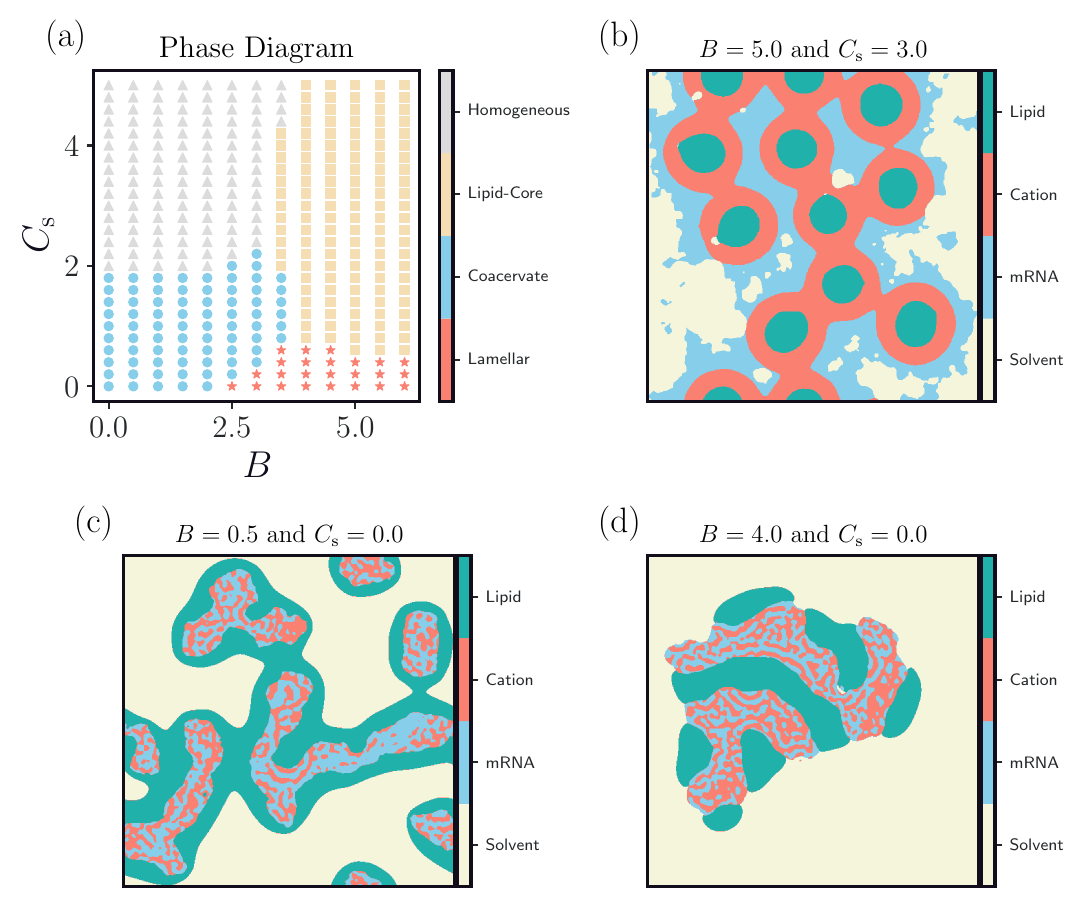}
    \caption{Morphological dependence on salt concentration and hydrophobicity (a) Phase diagrams with parameter cutoffs for  Homogeneous: $O_{\rm lipid} < 1.15$ \& $1 < O_{\rm ionic} < 1.0016$; Lipid-Core: $O_{\rm ionic} < 1.001$; Coacervate: $O_{\rm lipid} < 2.4$; Lamellar: All others. (b) Representative lipid-core phase (c) Coacervate-core phase (d)  Coacervate phase }
    \label{fig:phasediagram}
\end{figure}

A variety of experimental measurements, including dynamic light scattering (DLS)\cite{li_organ_2024} and $\zeta$-potential measurements~\cite{mckinlay_charge-altering_2017}, show that nanoparticles made with CARTs and similar polymeric materials form stable, predominantly spherical structures \cite{kaczmarek_polymerlipid_2016}.
These measurements, however, do not provide detailed insight into the microstructure (i.e., the spatial organization of the components within the nanoparticles). Understanding the connection between polymer properties and microstructure would be helpful in making predictions about nanoparticle stability.
As shown in Fig.~\ref{fig:phasediagram}, our model predicts a lamellar morphology for assembled nanoparticles in a range of conditions. 
Recent cryoEM data also support this viewpoint.

\begin{figure}[ht]
    \centering
    \includegraphics[width=\linewidth]{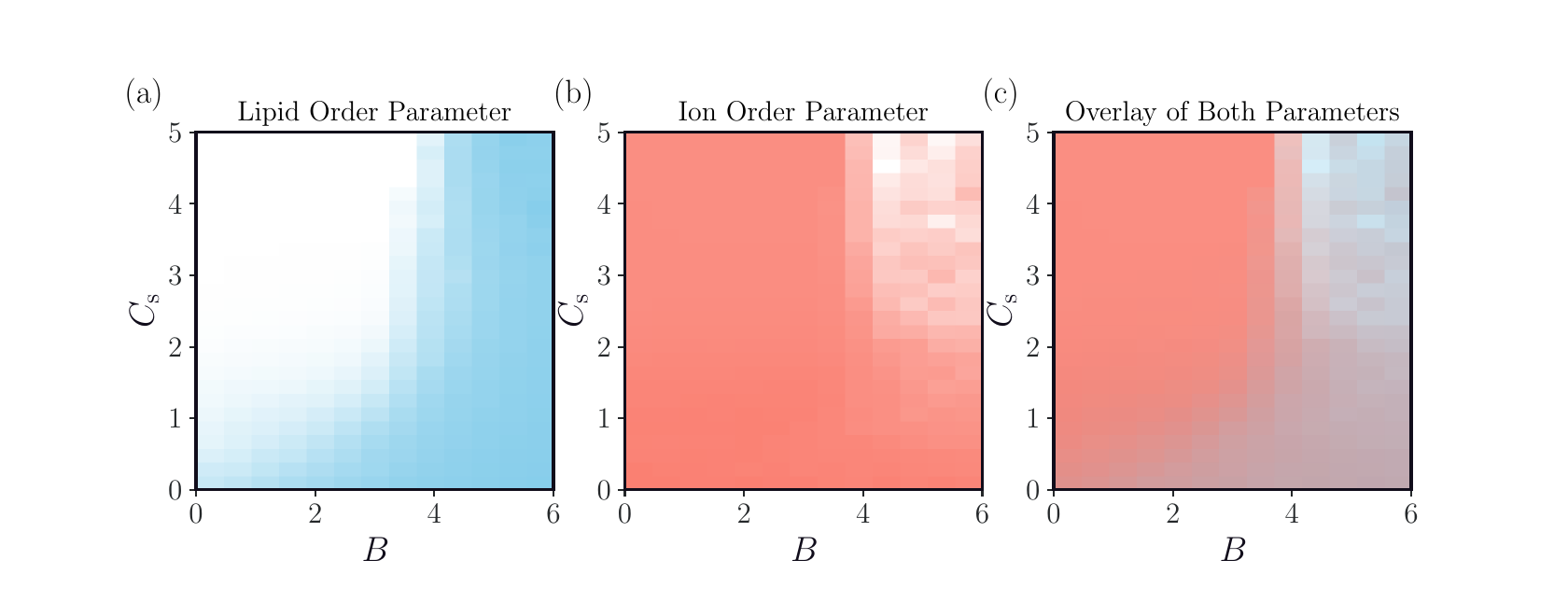}
    \caption{The two order parameters used to characterize the different morphologies. (a) Lipid density order paramter only (b) Charged species order parameter only (c) Overlay of both parameters at once}
    \label{fig:orderparams}
\end{figure}

\begin{figure}[ht]
    \centering
    \begin{subfigure}[t]{0.42\textwidth}
    \includegraphics[width=\linewidth]{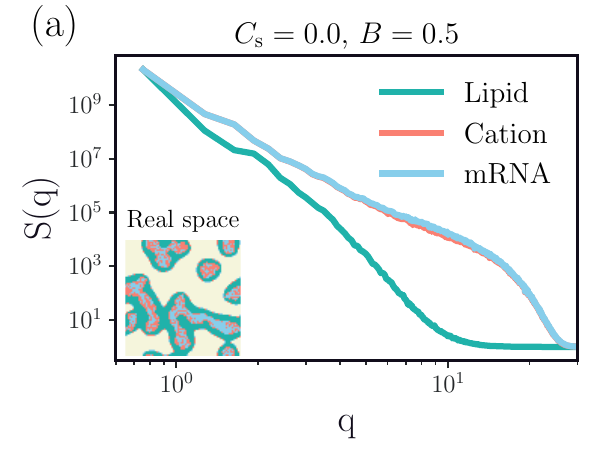}
        \label{fig:sub1}
    \end{subfigure}
    \hfill
    \begin{subfigure}[t]{0.42\textwidth}
        \includegraphics[width=\linewidth]{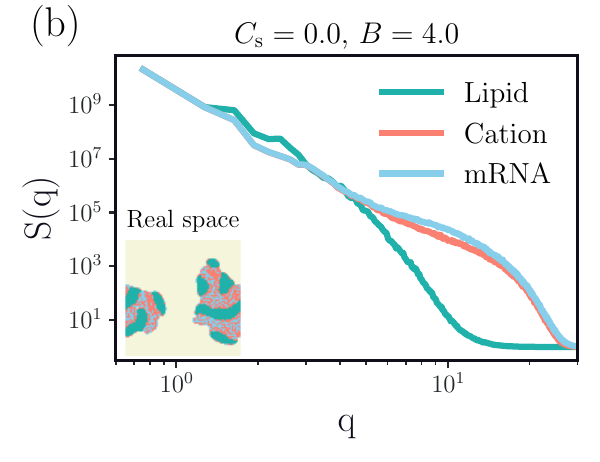}
        \label{fig:sub2}
    \end{subfigure}
    \begin{subfigure}[t]{0.42\textwidth}
        \includegraphics[width=\linewidth]{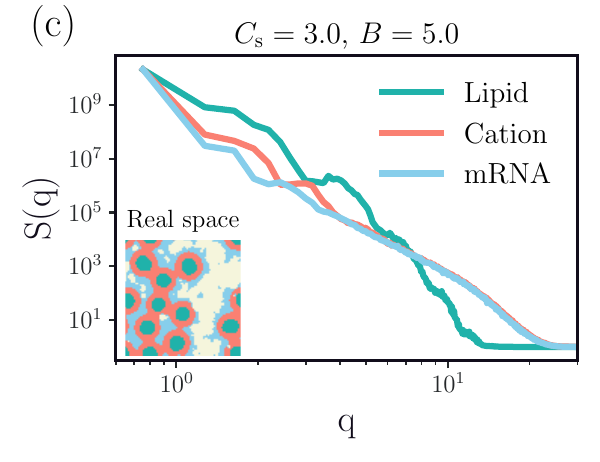}
        \label{fig:sub3}
    \end{subfigure}
    \caption{Comparison of structure factor for a representative member of each morphology. (a) Structure factor for the coacervate phase. (b) Structure factor for the lamellar phase (note higher correlation at low q values for lipid). (c) Structure factor for Lipid-Core phase.}
    \label{fig:struct}
\end{figure}

We choose to examine morphology along a range of salt concentrations and lipid hydrophobicity. Salt concentrations are experimentally variable and FH parameters are difficult to determine experimentally, so we choose to examine how behavior changes as they are varied. These two parameters essentially control the two possible driving forces for assembly (hydrophobicity and coacervation) so by changing them we can see how morphology is dependent on each effect. Examining the morphology at different points along these two axes we see 4 distinct morphologies in Fig. \ref{fig:phasediagram}. These were distinguished with two order parameters, 

\begin{equation}
\begin{aligned}
    O_{\rm lipid}(B, C_{\rm s}) = \frac{1}{V} \int_{\Omega} \rho_{\rm l} \frac{3 \rho_{\rm l} + \delta}{\rho_{\rm l} + \rho_+ + \rho_- + \delta} \d\rb, \\
    O_{\rm ionic}(B, C_{\rm s}) = \frac{1}{V} \int_{\Omega} \left(\rho_+ + \rho_-\right) \exp\left(-\frac{1 + \delta}{ |\rho_+ - \rho_- |+ \delta} \right) \d \rb. 
\end{aligned}
\end{equation}

These two parameters distinguish the separation between the lipid and charged species. The lipid parameter measures the relative density of the lipid compared to all species, and will get larger as the lipid aggregates. The ionic parameter measures the degree to which the positive and negative charges are separated. When it is larger they are all well paired, and as they become more mismatched it will shrink. Both are scaled by the local density of the species of interest to better correlate their effects with their representative species. We chose cutoffs such that the morphologies were distinguished as best as possible, though the transition is continuous, so the boundaries are somewhat arbitrary. We use $\delta = 0.0001$ to regularize and discard any points with a negative density (small negative densities are a consequence of complex Langevin sampling---the average will always be positive, but negative values may transiently occur). 

These parameters can be used to distinguish between four morphologies: one homogeneous and three condensed spatial arrangements that are connected by continuous transitions. 
The main difference between the separated phases is in the structure of the dense phase. 
In regions of high-salt and hydrophobic lipids the lipid segregation is the driving factor. Lipid segregation creates lipid-core micelles with a corona of the bound cation and a second corona of mRNA that is weakly attracted to the cationic block. In cases where the lipid is weakly hydrophobic and the salt concentration is low, we see phase separation driven by coacervation where the coacervate forms a dense phase with a corona of the bound lipid. For cases when there is low salt and hydrophobic lipid, both coacervation and hydrophobicity are driving forces, so we have a lamellar phase where both dense phases are in contact with the solvent. There is no coexistence between the lipid-core, lamellar, and coacervate phases because they require different conditions to exist, and they exist in a continuum as the parameters are varied. Even though smooth change between them is possible,  they still represent distinctive morphologies because they have varied structure factors, real densities, and driving forces.

We also computed the structure factor for each species, 
\begin{equation}
    S(\boldsymbol{q}) = \frac{1}{V} \int_V e^{-i \boldsymbol{q}\cdot (\rb-\rb')} \d \rb \d \rb',
\end{equation}
which provides insight into relevant length scales of features associated with each component. 
The structure factors, shown in Fig.~\ref{fig:struct}, evince differences between these three morphologies. The lipid core nanoparticle is the most distinctive, with a series of peaks well correlated with each of the three distinct nanoparticles morphologies. 
The spacing of these peaks depends on the micelle density, but the cation and mRNA have significantly decreased long-range correlation, which is characteristic of the phase. 
Both of the other two morphologies have significant long-range correlation of the charged species that is indicative of coacervation. The main distinguishing feature between the lamellar phase and the coacervate-core phase is the presence of a shorter range shoulder for the lipid-lipid correlation. This is only present in the lamellar phase because the coacervate core phase has a significantly less dense and organized lipid region. In interpreting these structure factors,  long range correlations are less reliable as they start to be effected by aberrations from the periodic box. 

\begin{figure}[h]
    \centering
    \begin{subfigure}[t]{0.4\textwidth}
        \centering
        \adjustbox{valign=t}{\includegraphics[width=\textwidth]{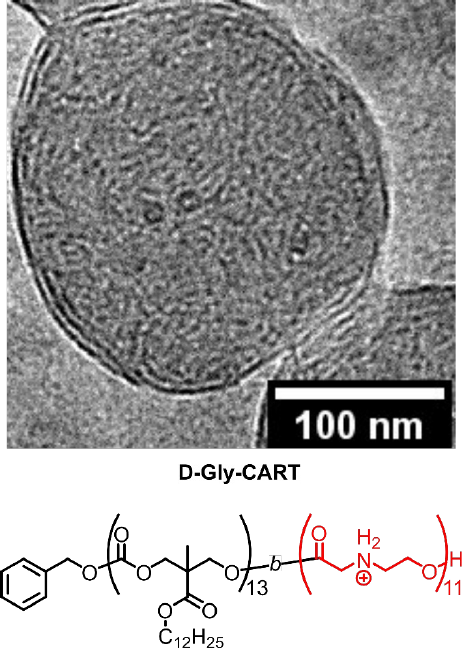}}
        \caption{Top: representative cryoEM micrograph of a CART/mRNA NP with a +/- ratio of 10/1. Below: the chemical structure of the CART, a dodecyl lipid carbonate block with a N-hydroxyethyl glycine-derived cationic block. }
        \label{fig:figure1}
    \end{subfigure}\hfill
    \begin{subfigure}[t]{0.4\textwidth}
        \centering
        \adjustbox{valign=t}{\includegraphics[width=\textwidth]{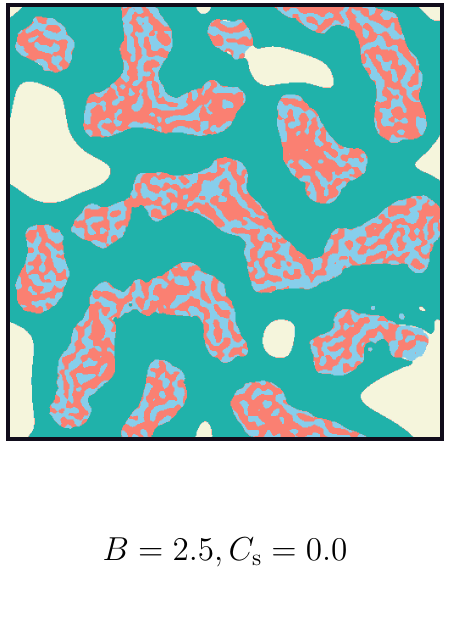}}
        \caption{Simulated CART-mRNA mixture with higher polymer density (63\% polymer by mass) at conditions corresponding to lamellar phase}
        \label{fig:figure2}
    \end{subfigure}
    \caption{Correspondence between unsalted CART in experiments and in the simulation.}
    \label{fig:sidebyside}
\end{figure}

We tested the predictions of our field theoretic simulations by conducting cryoEM imaging of CART-mRNA nanoparticles. 
We prepared a diblock copolymer with a combination of MTC-dodecyl carbonate for the hydrophobic block and N-hydroxyethyl glycine-derived 
\textalpha
-amino ester for the charged cationic block  
This diblock copolymer was combined with mRNA in a +/- charge ratio of 10:1 in 1 x PBS buffer at pH = 5.5 upon which rapid formation of nanoparticles was observed \cite{mckinlay_charge-altering_2017}.
The nanoparticles were then imaged using cryoEM to assess the organization within a nanoparticle. 
The structures shown in Fig.~\ref{fig:sidebyside} show clear lamellar ordering, a feature also observed in simulations with low salt and high hydrophobicity. 

There are a few notable differences between the experimental system and simulation due to simulation constraints. 
The size of a full nanoparticle is larger than the simulations that we can presently conduct.
Nevertheless, the mesoscale structure is well captured by our model. Experimentally, the mRNA is significantly longer than the cationic block of the CART polymer; however, numerical stability of our field theoretic simulation degrades as the individual chains become long. 
We did not observe significant morphological changes in the morphology for increasing mRNA block length in simulations stable enough to converge.
With this in mind, we chose the mRNA length to match the cationic block length, which provided high stability across conditions for the phase diagram Fig.~\ref{fig:phasediagram}.
Nevertheless, the model accurately captures clear lamellar structure in the CART-mRNA nanoparticles, resolving an open question about the internal arrangement of components in this model.

\section{Conclusions}

To determine the possible mesostructures for mRNA-CART nanoparticles, we built a complete and descriptive field theoretic model that incorporates hydrophobicity, explicit solvent and salt effects, and charge-charge interactions. 
The complexity of this model is a barrier to analytical treatment, so we relied on field theoretic simulation to compute a phase diagram that depends only on generic and controllable parameters of these materials.
Our results agree qualitatively with experimentally determined cryoEM, broadly showing lamellar mesoscale organization of nanoparticle contents. 

Because the driving forces balance coacervation and hydrophobicity, just as in biomolecular materials, including intrinsically disordered proteins, the toolkit developed here may be useful for studying so-called membraneless organelles~\cite{lyon_framework_2021}.
Furthermore, the lack of free and open-source tools for advanced simulation techniques using field theoretic approaches has been a barrier to adoption of these techniques~\cite{arora_broadly_2016}.
Our codebase is freely available and open-source and should serve to promote future work in field theoretic simulation.

\section{Methods: Experimental preparation of CARTs }

CART/mRNA nanoparticle preparation: A dodecyl-glycine CART was synthesized according to literature precedent \cite{mckinlay_charge-altering_2017}. CART/mRNA nanoparticles were formulated by mixing 17.72 $\mu\textrm{L}$ of a PBS buffer at pH = 5.5 with 1 $\mu\textrm{L}$ of a 1 mg/mL Fluc RNA solution (Trilink) and then adding 1.28 $\mu\textrm{L}$ of a stock solution of 2mM CART to the resulting solution to end up with a +/- charge ratio of 10/1. The solution was mixed for 20 s and immediately vitrified.

Cryogenic-electron microscopy (cryoEM) Grid Preparation and Imaging: 3 $\mu\textrm{L}$ of the CART/mRNA solution was applied to a Quantifoil R2/1 grid with 2 nm ultrathin carbon backing. Prior to loading, grids were glow discharged for 15 s at 10mA to increase hydrophilicity. Vitrification was carried out by Vitrobot (ThermoFisher). Grid preparation was performed at 100\% humidity and the grids were blotted for 3 s at a blot force setting of 2 prior to plunging into liquid ethane. CryoEM samples were clipped and then imaged on a ThermoFisher Glacios CryoEM. Images were recorded using SerialEM software in low dose mode with a K3 camera with an exposure time of 3 s and a defocus of -3 $\mu\textrm{m}$. 

\section{Methods: Numerical algorithms for complex Langevin simulation}

Scripts to run all numerical experiments in this paper are available at \url{https://github.com/rotskoff-group/polycomp.git}.
We benchmarked our code extensively against analytical results and previously published numerical results for coacervation that employ field theoretic simulation~\cite{delaney_theory_2017}.
Details of these benchmarks are given in Appendix~\ref{app:benchmark}.

Updating the fields and evaluating the free energy both require evaluations of the single-chain partition function, which is the most computationally difficult part of the Hamiltonian to evaluate \eqref{eq:MDE}. This equation requires evaluating both Laplacian and linear terms, so it is solved by a pseudospectral decomposition where the Laplacian update is done in reciprocal space while the other updates are done in real space. To improve the accuracy of this operation, we use a 4th order Richardson extrapolation scheme, which requires only 3 evaluations of each step. 

For the actual sampling, we use a complex Langevin scheme. This scheme builds trajectories in fictitious time in which the actual trajectory has no physical interpretation, but the statistical averages should converge to the correct values. Because field theoretic simulations aim to find a contour on the imaginary axis that has a local maximum for smooth integration, the sampling only happens in the real axis. Under this scheme, noise is injected into the real axis to sample it, while the imaginary axis relaxes to find local maxima that correspond to the states that are sampled over. 

Our actual system updates happen in Fourier space. This leads to a stiff system, so we follow previous work and implement an exponential time differencing (ETD) scheme to update the fields for the complex Langevin time stepping. The ETD scheme works by analytically solving the linear portion of a derivative to give an estimate of the relaxation rate of each mode. With this estimate in hand, we can essentially rescale the time step for each mode of our Fourier space representation. This allows all the modes to relax efficiently because their effective time step matches their relaxation rate, avoiding the need for the slower modes to relax at the same rate as the faster modes. Here we implement a first order ETD scheme, following previous work. Extensions to higher order ETD methods are possible, but not explored at this point\cite{cox_exponential_2002}. The mathematical details of all implementation details are available in the accompanying appendices. 

\section*{Acknowledgements}

Particular thanks is offered to Clay Batton, who provided invaluable help in understanding and debugging a number of algorithms in this paper and Sherry Li, whose skill and passion for vectorizing code and general understanding of mathematics was invaluable to this work. Additional appreciation to Jeremie Klinger for editing assistance. 
This material is based upon work supported by the U.S. Department of Energy, Office of Science, Office of Basic Energy Sciences, under Award Number DE-SC0022917.

\bibliographystyle{unsrtnat}
\bibliography{final-refs}

\clearpage
\newpage

\appendix

\section{Formal derivation of the field theoretic Hamiltonian} \label{app:deriv}

The material in this derivation is standard and builds on Refs.~\cite{fredrickson_equilibrium_2006, delaney_theory_2017, villet_efficient_2014}.
We first consider a system with three types of components: $P$ polymer species, $S$ solvent species and 2 salt species. The system has $M$ different types of monomers which make up the polymer and solvent. 
We rescale all lengths relative to a single reference polymer length $N$ and a reference bond stretching parameter $b$. 
For example, the volume $\Tilde{V}$ becomes $V = \Tilde{V} / {R_{\rm g}}^d$, where $R_{\rm g} = b/\sqrt{N/6}$ and $d$ is the system dimensionality. 
An unscaled position vector $\xb$ is rescaled in $V$ as $\rb := \xb / R_{\rm g}$.

Interactions between the monomers are governed by three distinct physical effects: FH surface interactions, Coulomb interactions for charged species, and bonded stretching interactions for polymer components. We model bonded interactions with a continuous Gaussian chain, where the stretching potential for a single polymer is a functional over a space curve representing the polymer
\begin{equation}
    U_0[\rb] = 
    \frac{3 k_{\rm B} T}{2} 
    \int_0^{N_p / N} \d s \vert \frac{\d \rb(s)}{\d s}\vert^2.
\end{equation}
Here, $s$ represents the scaled length variable indicating the parametric position along the polymer. 
Within this model, the partition function for an isolated chain can be computed analytically as a Gaussian functional integral.
We define the single chain partition function as 
\begin{equation}
    z_0 = \int_\Omega \mathcal{D}\rb e^{-\beta U_0[\rb]}.
\end{equation}
We compute this integral over all possible curves for the polymer type in question. 
For a non-interacting system of single chains,
\begin{equation}
    \mathcal{Z}_{\rm ideal} = \prod_i^P \frac{\left(z_{0_i} g_N^{(i)} V\right)^{n_i}}{n!} \prod_j^{S + 2} \frac{V}{\lambda_j^d N!},
\end{equation}
where $\lambda$ is the thermal wavelength for a given species and $g_N^{(i)}$ is the effective thermal wavelength for the Gaussian chain associated with species $i$. 

Before incorporating interactions, it is useful to define a density operator for any monomer species in our system. Each polymer of type $p$ has "structure" which we represent with a parameter $m_p(s)$ that indicates the monomer type at each point $s$ along $p$. 
Combining the density for both polymers and solvents the density operator is,
\begin{equation}
    \rho_m(\rb) \equiv 
    \int_\Omega \d \rb \sum_{p=1}^P \sum_{j=1}^{n_p}  \int_0^{N_j / N} \d s 
    \delta (\rb - \rb_{j}(s))\delta (m - m_p(s)) 
    + \sum_{s=1}^S \sum_{l=1}^{n_s} \delta (\rb 
    - \rb_{l})\delta (m - m_s).
\end{equation}

Because the interparticle interactions are formally divergent without regularization, we regularize the density and the charge density with a Gaussian kernel,
\begin{equation}
    \Gamma(\rb) = (2 \pi a^2)^{d/2} \exp \left(-\frac{1}{2 a^2} \rb^T \rb \right).
    \label{eq:gauss1}
\end{equation}
We also rescale the FH interactions $\chi_{ij}$ as 
$B_{ij} = \chi_{ij}N^2 / {R_{\rm g}}^d$. 
The Coulomb charge-charge interaction relies on the rescaled Bjerrum length $E = 4 Z^2 \pi N^2 \ell_{\rm B} / R_{\rm g}$. We simplify the parameter $E$ by setting the reference charge $Z=1$ and scale charges explicitly in the system if needed. 
Throughout, we assume that $E$ does not depend on the system composition due to the screening from the explicit salt in the system. 
Ignoring an arbitrary, finite constant, the charge-charge interaction is
\begin{equation}
    \beta U_{\rm C} = \frac{1}{2} \int_{\Omega} \d\rb\ \d \rb' \sum_i^M \sum_j^M  \Gamma * \rhob_i(\rb) 
    \frac{Z_i Z_j E}{|\rb - \rb'|} \Gamma * \rhob_j(\rb').
\end{equation}
This term simplifies if we define $\rho_{\rm C}(\rb) \equiv \sum_{i=0}^M Z_i \rho_i (\rb)$ to 
\begin{equation}
    \beta U_{\rm C} = \frac{1}{2} \int_{\Omega} \d\rb\ \d\rb'\ \Gamma * \rho_{\rm C}(\rb) 
    \frac{E}{|\rb - \rb'|} \Gamma * \rho_{\rm C}(\rb').
\end{equation}

\subsection{Flory-Huggins pair interaction}
After regularization and rescaling, the total Flory-Huggins interaction for a pair of species is
\begin{equation}
    \beta U_{ij} =\int_{\Omega} \d \rb\ \d \rb' B_{ij} \Gamma * \rhob_i (\rb)
    \delta(\rb-\rb')\Gamma * \rhob_j(\rb').
\end{equation}

A diagonal basis for the Flory-Huggins matrix yields the linear combinations of species that are physically distinct, and hence this basis is a natural one for developing a field theory. 
The FH matrix is given by
\begin{equation}
\chi = \begin{bmatrix} 
    2 \chi_{11} & \dots  & \chi_{1M}\\
    \vdots & \ddots & \vdots\\
    \chi_{M1} & \dots  & 2 \chi_{MM} 
    \end{bmatrix}.
\end{equation}
Because $\chi$ is a real, symmetric matrix, we can diagonalize it with the decomposition
\begin{equation}
    \rhob^T(\rb) \chi \rhob(\rb)
    = \left(\bb^T\rhob(\rb)\right)^T
    \textbf{B} \ \bb^T\rhob(\rb),
\end{equation}
where $\bb$ is the matrix of eigenvectors $\chi$ and 
$\textbf{B}$ is the corresponding diagonal matrix of eigenvalues. Additionally, we define a vector $\boldsymbol{\gamma}$ s.t. $\gamma_i$ takes value \textit{i} (value $\sqrt{-1}$) if $B_i > 0 $ and 1 otherwise. 
In this diagonal representation, 
\begin{equation}
    U_{\rm FH} = \frac{1}{2}\int_{\Omega} \d \rb\ \Gamma * \rhob^T(\rb)\chi \Gamma * \rhob(\rb) 
    = \frac{1}{2}\int_{\Omega} \d \rb \sum_{i=1}^M B_i \left(\bb^T \Gamma * \rhob(\rb)\right)^2_i.
\end{equation}
Conveniently, this representation also makes the partition function Gaussian: 
\begin{equation}
    \mathcal{Z} = \mathcal{Z}_{\rm ideal}\mathcal{Z}_{\rm C} \prod_{i=1}^M \int \mathcal{D} \left(\bb^T \rhob \right)_i e^{- \frac{\beta}{2} \int_\Omega \d\rb \int_\Omega \d\rb' \left(\bb^T \rhob(\rb)\right)_i B_i \delta(\rb - \rb') \left(\bb^T \rhob(\rb')\right)_i}.
\end{equation}
Here, $\mathcal{Z}_{\rm ideal}$ is the ideal gas partition function and $\mathcal{Z}_{\rm C}$ is the contribution from the charged part of the Hamiltonian which is discussed later.

We use a Hubbard-Stratonovich transform to integrate out the spatial density, which for a generic pair potential $u$ takes the form: 
\begin{equation}
    e^{-\frac{\gamma^2\beta}{2}\int_\Omega \d\rb \int_\Omega \d\rb' 
    \rho(\rb)u(\rb - \rb')
    \rho(\rb')}\\= 
    \frac{\int \mathcal{D} \omega e^{-\frac{1}{2\beta}\int_\Omega \d\rb
    \int_\Omega \d\rb'\omega(\rb)
    u^{-1}(\rb - \rb') \omega(\rb')
    - \gamma \int_\Omega \d\rb \omega(\rb)\rho(\rb)}}
    {\int \mathcal{D} \omega e^{-\frac{1}{2\beta}\int_\Omega \d\rb
    \int_\Omega \d\rb'\omega(\rb)
    u^{-1}(\rb - \rb') \omega(\rb')}}.
\end{equation}
For each of $B_i$ we carry out a Hubbard-Stratonovich transform, and making the substitution back into the partition function, we obtain
\begin{equation}   
    \mathcal{Z} = \mathcal{Z}_{\rm ideal}\mathcal{Z}_{\rm C} \prod_{i=1}^M \int \mathcal{D} \left(\bb^T \rhob \right)_i\frac{\int \mathcal{D} \omega_i e^{-\frac{1}{2\beta}\int_\Omega \d\rb
    \int_\Omega \d\rb'\omega_i(\rb)
    \frac{1}{B_i} \delta(\rb - \rb') \omega_i(\rb')
    - \gamma_i \int_\Omega \d\rb \omega_i(\rb)\left(\bb^T \Gamma * \rhob(\rb)\right)_i}}
    {\int \mathcal{D} \omega_i e^{-\frac{1}{2\beta}\int_\Omega \d\rb
    \int_\Omega \d\rb'\omega_i(\rb)
    \frac{1}{B_i} \delta(\rb - \rb') \omega_i(\rb')}}.
\end{equation}
Using the constraint imposed by the $\delta$-function and moving the convolution from the density to the conjugate field, which is valid because the convolution is only a function of $\rb$, we arrive at
\begin{equation}   
    \mathcal{Z} = \mathcal{Z}_{\rm ideal}\mathcal{Z}_{\rm C} \prod_{i=1}^M \int \mathcal{D} \left(\bb^T \rhob \right)_i \frac{\int \mathcal{D} \omega_i e^{-\frac{1}{2 B_i \beta}\int_\Omega \d\rb
    \omega_i(\rb)^2
    - \gamma_i \int_\Omega \d\rb \Gamma * \omega_i(\rb)\left(\bb^T \rhob(\rb)\right)_i}}
    {\int \mathcal{D} \omega_i e^{-\frac{1}{2 B_i \beta}\int_\Omega \d\rb
   \omega_i(\rb)^2}}.
\end{equation}
This equation can be further simplified by making a Wick rotation $\mu_i = \gamma_i \omega_i$ to give the final result, 
\begin{equation}   
    \mathcal{Z} = \mathcal{Z}_{\rm ideal}\mathcal{Z}_{\rm C} \prod_{i=1}^M \int \mathcal{D} \left(\bb^T \rhob \right)_i \frac{\int \mathcal{D} \mu_i e^{-\frac{\gamma^2_i}{2 B_i \beta}\int_\Omega \d\rb
    \mu_i(\rb)^2
    - \int_\Omega \d\rb \Gamma * \mu_i(\rb)\left(\bb^T \rhob(\rb)\right)_i}}
    {\int \mathcal{D} \mu_i e^{-\frac{\gamma^2_i}{2 B_i \beta}\int_\Omega \d\rb
   \mu_i(\rb)^2}}.
\end{equation}

\subsection{Charge-charge Interactions}

We now compute the contribution to the partition function from charge-charge interactions,
\begin{equation}
    \mathcal{Z}_{\rm C} = \int \mathcal{D} \rho_\textrm{C} e^{\beta U_{\rm C}} =  \int \mathcal{D} \rho_\textrm{C} e^{\frac{1}{2} \int_{\Omega} \d\rb
    \int_{\Omega} \d\rb' \Gamma * \rho_{\rm C}(\rb) 
    \frac{E}{|\rb - \rb'|} \Gamma * \rho_{\rm C}(\rb')}.
\end{equation}
We again use a Hubbard-Stratonovich transform, noting that the functional inverse of $\frac{E}{|\rb - \rb'|}$ is $\delta(\rb - \rb')\frac{\nabla^2}{E}$, to give 
\begin{equation}   
    \mathcal{Z}_{\rm C} = \int \mathcal{D} \rho_\textrm{C} 
    \frac{\int \mathcal{D} \varphi e^{-\frac{1}{2E}\int_\Omega \d\rb
    \int_\Omega \d\rb'\varphi(\rb)
    \delta(\rb - \rb')\nabla^2 \varphi(\rb')
    - i \int_\Omega \d\rb \varphi(\rb) \Gamma * \rho_{\rm C}(\rb)}}
    {\int \mathcal{D} \varphi e^{-\frac{1}{2E}\int_\Omega \d\rb
    \int_\Omega \d\rb'\varphi(\rb)
    \delta(\rb - \rb')\nabla^2 \varphi(\rb')}}.
\end{equation}
We contract over the delta function, make a Wick rotation, and move the convolution to obtain
\begin{equation}   
    \mathcal{Z}_{\rm C} = \int \mathcal{D} \rho_\textrm{C} 
    \frac{\int \mathcal{D} \varphi e^{-\frac{1}{2E}\int_\Omega \d\rb
    \nabla^2 \varphi(\rb)^2
    - i \int_\Omega \d\rb \Gamma * \varphi(\rb) \rho_{\rm C}(\rb)}}
    {\int \mathcal{D} \varphi e^{-\frac{1}{2E}\int_\Omega \d\rb
    \nabla^2 \varphi(\rb)^2}}.
\end{equation}
Hence, the partition function involves an integral over the field $\varphi$. We can take the integral over $\rhob$ and $\rho_\textrm{C}$ to simplify our equation and give us the single chain partition function
\begin{equation}
    \mathcal{Z} = \mathcal{Z}_{\rm ideal}  \mathcal{Z}_{\rm C} \prod_{i=1}^M \mathcal{Z}_i = 
    \mathcal{Z}_{\rm ideal} \int \mathcal{D} \varphi \prod_{i=1}^M \int \mathcal{D} \mu_i 
    e^{-H[\{\mu_i\}, \varphi]}
\end{equation}
with the field theoretic Hamiltonian
\begin{equation}
    H[\{\mu_i\}, \varphi] = 
    \sum_{i=1}^M \frac{\gamma_i^2}{2 B_i} 
    \int_\Omega \d\rb \mu_i^2(\rb)\\
    + \frac{1}{2 E} \int_\Omega \d \rb | \nabla \varphi(\rb)|^2\\
    - \sum_{j=1}^{P+S+2} n_j \log Q_j[\{\mu_i\}, \varphi].
\end{equation}

The equations we obtained are written in the diagonal basis of the chemical potential fields $\mu_i$, each of which couples to a linear combination of polymer species.
In the basis of the enumerated species, 
\begin{equation}
    \boldsymbol{\psi}(\rb) = \bb \boldsymbol{\mu}(\rb) + \boldsymbol{Z} \varphi(\rb)
\end{equation}
Our model assumes no substantive FH interactions among the salt species, so $\psi_{S} = Z_{\rm S} \varphi(\rb)$. 

The single particle partition function for species $j$ is
\begin{equation}
    Q_j = \frac{1}{V} \int_\Omega \d\rb e^{-\Gamma * \psi_j(\rb) / N}
\end{equation}
with corresponding density 
\begin{equation}
    \rho_j(\rb) = \frac{\partial \log Q_j}{\partial \psi_j(\rb)} = \frac{C_j}{Q_j} e^{-\Gamma * \psi_j(\rb) / N}
\end{equation}
where concentration $C_j := n_j / V$.

For the polymer species, we use the single \emph{chain} partition function,
\begin{equation}
    Q_j = \frac{1}{V} \int_\Omega \d\rb q_j(\frac{N_j}{N}, \rb)
\end{equation}
where $q_j$ evolves according to the modified diffusion equation
\begin{equation}
    \frac{\partial q_j(s, \rb)}{\partial s} = 
    \nabla^2 q_j(s, \rb) - \psi_j(s, \rb) 
    q_j(s, \rb) \label{eq:MDE}
\end{equation}
and $\psi_j(s, \rb)$ is position-dependent field corresponding to the monomer type parametrically at position $s$ along the curve of the polymer. 
We solve the modified diffusion equation with the following initial condition:
\begin{equation}
    q_j(0, \rb) = 1.
\end{equation}
The corresponding density is 
\begin{equation}
    \rho_m(\rb) = \frac{C_j}{Q_j}\int_0^{N_j/N} \d s q_j(s, \rb) q_j^{\dagger}(s, \rb) \delta (m - m(s)). 
\end{equation}
The adjoint $q_j^{\dagger}(s, \rb)$ is defined analogously to $q_j(s, \rb)$, though starting at the opposite polymer end.
That is, $q_j^{\dagger}(s, \rb)$ is the solution of 
\begin{equation}
    -\frac{\partial q_j^{\dagger}(s, \rb)}{\partial s} = 
    \nabla^2 q_j^{\dagger}(s, \rb) - \psi_j(s, \rb) 
    q_j^{\dagger}(s, \rb) \label{eq:MDEr};
    \qquad 
    q_j^\dagger(\frac{N_j}{N}, \rb) = 1.
\end{equation}
If a monomer type exists in multiple polymers or solvents then we can simply calculate the amount each species contributes to that monomer's density and sum to get the total density.

Previous derivations finished here, but we can analytically solve the homogeneous part of the free energy to avoid having to iteratively update the average value of the fields. 
The average field values do not effect many observables (such as density) but can effect other such as free energy and chemical potential. 
While the homogeneous solution can be approximated during the complex Langevin update scheme, we opt to instead solve exactly the homogeneous component and set the average field value for all fields to zero.
Conveniently because of charge conservation $\<\varphi\> = 0$, so there is no electrostatic free energy for the homogeneous case and no need to add any analytic term.
The homogeneous solution does contribute to the free energy for the FH terms, so we define a new vector $\cb = \frac{1}{V}\int_\Omega \d \rb \rhob$.
We can use the normal definition of FH interaction and the trivial solution for the homogeneous case to add the proper analytic correction.
Using this correction we can constrain all the fields s.t. $\<\mu_i\> = 0$, the Hamiltonian becomes 
\begin{equation}
    H[\{\mu_i\}, \varphi] = 
    \sum_{i=1}^M \frac{\gamma_i^2}{2 B_i} 
    \int_\Omega \d\rb \mu_i^2(\rb)\\
    + \frac{1}{2 E} \int_\Omega \d \rb | \nabla \varphi(\rb)|^2\\
    - \sum_{j=1}^{P+S+2} n_j \log Q_j[\{\mu_i\}, \varphi] + \frac{V \cb^T \chi \cb}{2}.
\end{equation}

\section{Numerical evaluation of the field theoretic Hamiltonian}

Equilibrating the system requires repeated computation of the density of all species, which involves solving the pair of modified diffusion equations \eqref{eq:MDE} and \eqref{eq:MDEr}. 
Efficient numerical schemes for such PDEs are both well-studied and widely used~\cite{villet_efficient_2014}.
We employ standard integration schemes using a splitting scheme together with the pseudospectral method. 

The splitting scheme we use alternates updates of the linear part of the MDE with expensive evaluations of the Laplacian operator, 
 \begin{equation}
      q(s + \Delta s, \rb) = e^{-\frac{\psi(s, \rb)}{2} \Delta s} e^{\nabla^2 \Delta s}e^{-\frac{\psi(s, \rb)}{2} \Delta s} q(s, \rb) + O(\Delta s^3). 
 \end{equation}
The operator $e^{\nabla^2 \Delta s}$ is calculated in Fourier space, which is the diagonal basis for the Laplacian.
We use a method based on Richardson extrapolation~\cite{ranjan_linear_2008} that has fourth order error. 
We calculate two estimates of $q_j(s + \Delta s, \rb)$, one with a single step of $\Delta s$ and one with two steps of $\Delta s / 2$ and use the extrapolation
\begin{equation}
    q(s + \Delta s, \rb) =  \frac{4 q_{\Delta s / 2}(s + \Delta s, \rb) - q_{\Delta s}(s + \Delta s, \rb)}{3}
\end{equation}
to obtain an approximation that is $O(\Delta s^4)$ with three evaluations of the MDE.

\section{Field theoretic simulation}\label{app:CL}

Because coacervation arises due to fluctuations, mean-field approximations do not capture the underlying physics accurately. 
The mean-field solution for a soluble polyampholyte is always a homogeneous charge neutral solution, inconsistent with the phenomenology of coacervation.
Monte Carlo methods and complex Langevin are both options for sampling fluctuations in the local density of each species.
Building on many previous successful applications~\cite{villet_efficient_2014, delaney_theory_2017, mckinlay_charge-altering_2017}, we have opted to use complex Langevin.

While it is well-known that complex Langevin lacks rigorous theoretical foundations and strong convergence guarantees, numerical results in the polymer field literature are in good agreement with both analytical and experimental results. 
In principle, the complex Langevin algorithm is relatively straightforward, essentially coupling gradient descent to a noise term. 
The main difference is that for complex Langevin, all of the chemical potential fields are promoted to being fully complex, allowing sampling and descent over both parts of the plane. 
The noise injected during sampling can be either complex or real, but in this case the noise will all be purely real before Wick rotation. 
Previous work \cite{fredrickson_equilibrium_2006}, has shown that any choice of noise should correctly sample the distribution, but the choice to only use real noise has a physical interpretation. For each field, sampling over the real axis is the goal, but it is generally easiest to do so by finding a saddle point at a different point in imaginary space and use Cauchy's theorem to verify that this gives the same result as the real integral. As such, it is natural to expect that the real goal is to sample the real axis, and condition this sampling on finding the corresponding maximum value of the imaginary axis. This regime corresponds directly to injecting noise only into the real part of the field. This approach is known as the "standard" complex Langevin and has been chosen here consistent with previous work \cite{fredrickson_equilibrium_2006}. With that in mind, we will write down the general form of the complex Langevin where we have promoted $({\omega_i}, \varphi)$ to be complex fields like $\zb = \textbf{w} + i \textbf{v}$  where \textbf{w} and \textbf{v} are both purely real. 

\begin{eqnarray}
    \frac{\partial}{\partial t} \textbf{w}(t) &=& -\lambda \textrm{Re}\left[\frac{\partial H(\zb(t))}{\partial \zb}\right]  + \boldsymbol{\eta}(t)\\
    \frac{\partial}{\partial t} \textbf{v}(t) &=& -\lambda \textrm{Im}\left[\frac{\partial H(\zb(t))}{\partial \zb}\right]. 
\end{eqnarray}

$t$ is a fictitious time variable constructed for sampling purposes and $\lambda$ is a time-step scale. In practice the actual equations are a series of vector equations, where we can write the force on all fields as 

\begin{equation}
    \textbf{F} (\boldsymbol{\mu}(t, \rb)) = 
    \frac{\partial H[\boldsymbol{\mu}(\rb), \varphi(\rb)]}{\partial \boldsymbol{\mu}(t, \rb)} 
    = \left(\frac{\boldsymbol{\gamma}^2}{\textbf{B}} \right)
     \odot \boldsymbol\mu (t, \rb)- 
    \bb^T\boldsymbol{\rho}(t, \rb) 
\end{equation}
and 
\begin{equation}
    F (\varphi(t, \rb)) = 
    \frac{\partial H[\boldsymbol{\mu}(\rb), \varphi(\rb)]}{\partial \varphi(t, \rb)} 
    = \frac{1}{E} \nabla^2 \varphi(\rb) - 
    \rho_{\rm C} (\rb).
\end{equation}

where $\odot$ represents the Hadamard product. We can also define the Fourier transformed forces, which avoid calculating convolutions and gradients directly as 
\begin{equation}
    \hat{\textbf{F}} (\boldsymbol{\mu}(t, \kb)) =  
    \mathscr{F}\left(
    \frac{\partial H[\boldsymbol{\mu}(\rb), \varphi(\rb)]}
    {\partial \boldsymbol{\mu}(t, \rb)} \right)
    = \left(\frac{\boldsymbol{\gamma}^2}{\textbf{B}} \right)
     \odot \hat{\boldsymbol\mu} (t, \kb)- 
    \bb^T\hat{\boldsymbol{\rho}}(t, \kb) 
\end{equation}
and
\begin{equation}
    \hat{\textbf{F}} (\varphi(t, \kb)) = \mathscr{F}\left(
    \frac{\partial H[\boldsymbol{\mu}(\rb), \varphi(\rb)]}{\partial \varphi(t, \rb)} \right)
    = \frac{1}{E} \kb^2 \hat{\varphi}(\kb) - 
    \hat\rho_{\rm C} (\kb). 
\end{equation} 

We can finally write down the update step for a first-order Euler–Maruyama as  
\begin{equation}
    \frac{\partial \boldsymbol{\mu}(t, \kb)}{\partial t} = - \boldsymbol{\lambda}_\mu \hat {\textbf{F}}(\boldsymbol{\mu}(t, \kb)) + \boldsymbol\gamma \odot \hat{\boldsymbol{\eta}}_{\mu}(t, \kb)
\end{equation}
\begin{equation}
    \frac{\partial \varphi(t, \kb)}{\partial t} = - \lambda_\varphi \hat {\textbf{F}}(\varphi(t, \kb)) + i \hat{\eta}_{\varphi}(t, \kb). 
\end{equation}

Because some of our fields have been Wick rotated, their corresponding noise is rotated in the same way. The noise terms are Fourier transforms of Guassian white noise with the following statistics 

\begin{eqnarray}
    \langle\eta_{i\mu}(t, \rb)\rangle = \langle\eta_{\varphi}(t, \rb)\rangle &=& 0\\
    \langle\eta_{i\mu}(t, \rb) \eta_{i'\mu}(t', \rb')\rangle &=&
    \frac{2 \lambda_{i\mu} \beta_{i\mu} }{\Delta V } \delta(i-i') \delta(t-t') \delta(\rb-\rb')\\
    \langle\eta_{\varphi}(t, \rb) \eta_{\varphi}(t', \rb')\rangle &=&
    \frac{2 \lambda_{\varphi} \beta_{\varphi} }{\Delta V } \delta(t-t') \delta(\rb-\rb').
\end{eqnarray}

$\beta_{i\mu}$ and $\beta_{\varphi}$ are temperatures corresponding to each field. They are not without physical consequence on the distribution reached, especially in the case of coacervation. 

In practice, the system in the Fourier basis is very stiff and EM1 or any Runge-Kutta method usually requires prohibitively small time steps to be practical. Previous studies \cite{villet_efficient_2014} have examined other integration schemes, and exponential time differencing (ETD) has generally proven to be the preferred method for integrating these equations. The general form for ETD is 

\begin{equation}
    \hat{w}(t + \Delta t) = \hat{w}(t)
    -\frac{1 - e^{-\lambda \Delta t c(\kb)}}{c(\kb)} \hat F(\textbf{w}(t)) + \left(\frac{1 - e^{-2 \lambda \Delta t c(\kb)}}{2 \lambda \Delta t c(\kb)} \right)^2 \hat\eta(t)
\end{equation}

where 

\begin{equation}
    c(\kb) \equiv \frac{\partial \hat F(\kb)}{\partial \hat w(\kb)}\bigg|_{\hat w(\kb) = 0}.
\end{equation}
This equation uses all the same variables as the EM scheme listed above. $c(\kb)$ is somewhat troublesome as it requires an analytical approximation to properly rescale the relaxation rates for each field. In practice we only need to know the partial differential between density and fields, because the other components of force are trivially solvable. The full derivation is a tedious extension of previous work \cite{liu_weak_2019} but the final result is 

\begin{gather}
    \frac{\partial \rho(\kb, [\hat \mu])}{\partial \hat{\mu}_i(\kb)}  \bigg|_{\{\hat \mu(\kb)\} = 0} =  \gamma_i^2 C \sum_j \hat g_{ij} \\
    \hat g_{ii} = - \left( f_{i+1} - f_i\right)^2 g_D\left(\left( f_{i+1} - f_i\right) k^2\right)\\
    \hat g_{ij} =\frac{1}{ k^4}\biggl( e^{-|f_{i+1} - f_{j+1}|k^2} - e^{-|f_{i+1} - f_j|k^2} + e^{- |f_{i} -  f_{j}|k^2} - e^{-  |f_{i} -  f_{j+1}|k^2} \biggr)
\end{gather}
where $\hat g_D$ is the Debye function 
\begin{equation}
    \hat g_D(k^2) = \frac{2}{k^4} \left( e^{-k^2} + k^2 - 1\right). 
\end{equation}
The values $\{f_0, f_1, ..., f_{n+1}\}$ are the set of break points along a polymer with $n$ blocks where a block of one monomer type gives way to another. By convention $f_0 =0$ and $f_{n+1} = N$ for all polymers. 
\section{Soft explicit solvent}

One novel issue that arises in this simulation is the requirement to have attractive FH interactions and coacervation in the same model. Previous publications have only included charge-charge interactions and repulsive FH interactions and used an implicit solvent\cite{delaney_theory_2017}. The other method that is commonly used for repulsive interactions is a hard explicit repulsion which treats the field associated with the total density having a different form that corresponds to infinite FH parameter \cite{duchs_multi-species_2014}. This acts as a hard enforcement of the total density constraint where there is no penalty for the value of the chemical potential field corresponding to total density. The implicit solvent model is not tenable here because it only works when all FH interactions are repulsive after diagonalization. The hard explicit solvent model is in theory not prohibitive, but is generally impractical because the hard constraints on total density make the system stiffer. The dynamic coacervation causes large fluctuations in total density, which lead to unstable dynamics when coupled with a scheme that tries to rigorously enforce total density. 

One solution that has been proposed and there are indications works favorably is the soft-explicit model  which has been described here. As far as we know there are no published results of the soft explicit model save a brief positive description \cite{villet_advanced_2012}. The soft explicit model is an extension of the implicit solvent model where all fields are the same but explicit solvent particles are added. This can be interpreted in two ways. First, it can be interpreted as an explicit solvent model where the total density is only weakly enforced, with the degree of enforcement being controlled by the self-interaction FH parameter ($B_{ii})$. In the limit of all $B_{ii} \rightarrow \infty$, there is no penalty on the field of the total volume, and the hard solvent behavior is reestablished. The second interpretation is that the simulation now includes one additional solvent species, an extra implicit co-solvent that fills the extra space left in the uneven density. In this case the $B_{ii} \rightarrow \infty$ limit can be interpreted as the implicit solvent being infinitely repulsive to all other species and thus enforcing total density. Both interpretations are reasonable and are useful in interpreting behavior depending on the question to be evaluated. In either case the soft explicit solvent gives the ability to tune the enforcement of the total density constraint that allows for simulation of dynamic systems.

\section{Comparison with Delaney and Fredrickson}\label{app:benchmark}
To validate our implementation, we made direct comparisons with both previous simulations of coacervation in implicit solvent and analytical results~\cite{delaney_theory_2017}. 
To make these comparisons, we implemented a chemical potential operator, an osmotic pressure operator, and the Gibbs ensemble. 
These operators are described in detail here before the direct comparisons are presented with previous results.

\subsection{Free Energy}

The free energy is has essentially already been defined, but we will explicitly express the ideal gas term so that we can take proper derivatives even between different system states. Doing so gives us the free energy functional 
\begin{equation}
    F(T) = kT\biggr(\sum_j^{P + S + 2} \left(-n_j\log z_{j0} - n_j \log V + n_j \log n_j - n_j \right) 
    + \<H[\{\mu_i\}, \varphi]\>_{\rm S}\biggr)
\end{equation}
\begin{equation}
    F(T) = kT \left(\sum_j \left(-n_j\log z_{j0} + n_j \log C_j - n_j \right) + \<H[\{\mu_i\}, \varphi]\>_{\rm S}\right)
\end{equation}
Here the average $\<H[\{\mu_i\}, \varphi]\>_{\rm S}$ is the average over a well equilibrated simulation that approximates the infinite functional integral in question. 

\subsection{Chemical potential operator}
The chemical operator is defined for each species of polymer, solvent and salt as
\begin{equation}
    \mu_j = \frac{\partial F(T)}{\partial n_j}
\end{equation}
which requires solving 
\begin{equation}
    \frac{\partial H}{\partial n_j} = \frac{\partial}{\partial
 n_j}\left(
    \sum_{i=1}^M \frac{\gamma_i^2}{2 B_i} 
    \int_\Omega \d\rb \mu_i^2(\rb)\\
    + \frac{1}{2 E} \int_\Omega \d \rb | \nabla \varphi(\rb)|^2\\
    - \sum_{j=1}^{P+S+2} n_j \log Q_j[\{\mu_i\}, \varphi] + \frac{V \cb^T \chi \cb}{2}\right). 
\end{equation}
The first two terms do not depend on this derivative at all, and the single species partition is trivial. The remaining analytic term is the only one that poses any difficulty. This is mainly because the vectors in that term are indexed via monomer type rather than species, so we have to make the expansion
\begin{equation}
    \frac{\partial \textbf{c}}{\partial n_j} =
    \begin{bmatrix} 
    \frac{\partial C_1}{\partial n_j} \\
    \vdots \\
    \frac{\partial C_i}{\partial n_j} \\
    \vdots \\
    \frac{\partial C_M}{\partial n_j} 
    \end{bmatrix}
    = \frac{1}{V}\begin{bmatrix} 
    \frac{\partial n_1}{\partial n_j} \\
    \vdots \\
    \frac{\partial n_i}{\partial n_j} \\
    \vdots \\
    \frac{\partial n_M}{\partial n_j} 
    \end{bmatrix}
    = \frac{1}{V} \begin{bmatrix} 
    \frac{\kappa_{j1}N_j}{N}  \\
    \vdots \\
    \frac{\kappa_{ji}N_j}{N} \\
    \vdots \\
    \frac{\kappa_{jM}N_j}{N} 
    \end{bmatrix}
    = \frac{\vec \kappa}{V}
\end{equation}
where $\kappa_{ji}$ is the fraction of the $j$-th species that is comprised of the $i$-th monomer type. If the $j$-th species is a solvent or a homopolymer with monomer type $i$ then, $\kappa_{ji} = 1$ and all other $\kappa_{ji'}=0$. Similarly, if the $j$-th species is a diblock copolymer with equal fraction of species 1 and 2, then $\kappa_{j1} = \kappa_{j2} = 1/2$, with all other $\kappa_{ji'}=0$.

Now we can compute,
\begin{equation}
    \frac{\partial}{\partial n_j}\frac{V \textbf{c}^T \chi \textbf{c}}{2} 
    = \frac{V \vec \kappa^T \chi \textbf{c}}{2V} + \frac{V \textbf{c}^T \chi \vec \kappa}{2V}
\end{equation}
where because $\chi$ is symmetric, $\vec \kappa^T \chi \textbf{c} = \textbf{c}^T \chi \vec \kappa$, and 
\begin{equation}
    \frac{\partial}{\partial n_j}\frac{V \textbf{c}^T \chi \textbf{c}}{2} 
    = \vec \kappa^T \chi \textbf{c}.
\end{equation}

With this we can write down the total free energy, where the only other terms are easy analytic derivatives of the free energy. 

\begin{equation}
    \mu_j = \frac{\partial F(T)}{\partial n_j} = kT \left(-\log z_{j0} + \log C_j  - \<\log Q_j[\{\mu_i\}, \varphi]\>_{\rm S} + \vec \kappa^T \chi \textbf{c} \right)
\end{equation}

In practice the $\log z_{j0}$ term is always the same between different simulations and will never contribute in a meaningful way to comparing states so it will be ignored. 

\subsection{Pressure Operator}

Another operator that we will need to equilibrate our system is the osmotic pressure which represents changes in the amount of volume in the system (and thus implicit solvent). This operator is generically defined as 

\begin{equation}
    \Pi = -\frac{\partial F(T)}{\partial V}. 
\end{equation}

This operator provides multiple complexities, as we shall soon see, but we can start with the relatively easier term associated with the homogenous field. We will start by noting that $\frac{\partial C_j}{\partial V} = -C_j/V$ then proceed to 
\begin{equation}
    \frac{\partial}{\partial V} \frac{V \textbf{c}^T \chi \textbf{c}}{2} = \frac{\textbf{c}^T \chi \textbf{c}}{2} - \textbf{c}^T \chi \textbf{c} = -\frac{\textbf{c}^T \chi \textbf{c}}{2}. 
\end{equation}

\subsubsection{Hamiltonian Derivative}

The Hamiltonian derivative is significantly more complicated. Most of this derivation will closely follow a previous derivation \cite{villet_efficient_2014} with modifications to generalize across system dimensionality, polymer structure, and charge. We will ignore the FH contribution term as we have already handled it and it does not have to be sampled over. For everything else, we note that 

\begin{align}
    \beta \Pi_{ex} &= -\frac{\partial}{\partial V} \log \frac{\mathcal{Z}_{\rm C}}{\mathcal{Z}_0} \nonumber \\
    &= -\frac{\partial}{\partial V} \log 
    \frac{\prod_i\int\mathcal{D}\mu_i\int \mathcal{D} \varphi 
     e^{-H[\{\mu_i\}, \varphi]}}{\prod_i\int\mathcal{D}\mu_i 
    e^{-\frac{\gamma_i^2}{2 B_i} 
    \int_\Omega d\rb\mu_i(\rb)^2}\int \mathcal{D} \varphi 
    e^{-\frac{1}{2E}\int_\Omega d\rb| \nabla \varphi(\rb)|^2}}. 
\end{align}

As in the previous derivation, we need to rescale all $\{\mu_i\}, \varphi$, such that the $\frac{\partial }{\partial V}$ with respect to the denominator vanishes. It was previously shown that the correct rescaling for $\mu$ is $\mu(\rb) = V^{-\frac{1}{2}} w(\rb)$ and we will posit and demonstrate that the correct rescaling for $\varphi$ is $\varphi(\rb) = V^{-1/2d}f(\rb)$, where $d$ is the system dimensionality.  We will now show that this is the correct rescaling by splitting the logarithm and taking each individual derivative

\begin{equation}
\begin{aligned}
    \beta \Pi_{ex} &=& \frac{\partial}{\partial V} \biggl(-\log\bigl( \prod_i\int\mathcal{D}w_i\int \mathcal{D} f 
    e^{-H[\{V^{-1/2}w_i\}, V^{-1/2d} f]}\bigl) \\
    &+& \sum_i \log (\int\mathcal{D}w_i 
    e^{-\frac{\gamma_i^2}{2 B_i } 
    \int_\Omega d\rb(V^{-1/2}w_i(\rb))^2})\\
    &+& \log\int \mathcal{D} f 
    e^{-\frac{1}{2E}\int_\Omega d \rb
    | \nabla (V^{-1/2d}f(\rb))|^2}\biggr)\\
    &=& \frac{\prod_i\int\mathcal{D}w_i\int \mathcal{D} f 
    \frac{\partial H[\{V^{-1/2}w_i\}, V^{-1/2d} f]}{\partial V}e^{-H[\{V^{-1/2}w_i\}, V^{-1/2d} f]}}
    {\prod_i\int\mathcal{D}w_i\int \mathcal{D} f 
    e^{-H[\{V^{-1/2}w_i\}, V^{-1/2d} f]}} \\
    &+& \frac{\partial}{\partial V}\sum_i \log \int \mathcal{D} w_i e^{-\frac{\gamma_i^2}{2 B_i V}
    \int_\Omega d\rb w_i(\rb)^2}\\ 
    &+&  \frac{\partial}{\partial V} \log\int \mathcal{D} f 
    e^{-\frac{1}{2E}\int_\Omega d \rb
    | \nabla (V^{-1/2d} f(\rb))|^2}\biggr).
\end{aligned}
\end{equation}

Both of the last two terms are identically zero. They are both Gaussian integrals that have no V dependence. While this is more difficult to see for the $\varphi$ term, but if we use the convention that $\zb = V^{-1/d} \rb$, we can do the intermediate derivative explicitly, which is 
\begin{equation}
\begin{aligned}
    \frac{\partial}{\partial V} | \nabla (V^{-\frac{1}{2d}} f(\rb))|^2 &= \frac{\partial}{\partial V} \left(\sum_i^{d} V^{\frac{1}{d}}\frac{\partial}{\partial \zb_i} V^{-\frac{1}{2d}}f(\rb)\right)^2 \\
    &=  2 \nabla (V^{-\frac{1}{2d}} f(\rb))  \left(\sum_i^{d} \frac{\partial}{\partial \zb_i} f(\rb)\right)\frac{\partial}{\partial V} V^{\frac{1}{2d}} \\
    &=  2 \nabla (V^{-\frac{1}{2d}} f(\rb)) \left(\sum_i^{d} \frac{\partial}{\partial \zb_i} f(\rb)\right) \frac{1}{2d}V^{\frac{1}{2d} - 1} \\
    &=\frac{1}{Vd}| \nabla (V^{-\frac{1}{2d}} f(\rb))|^2. 
\end{aligned}
\end{equation}

If we propagate this through, the extra term of $1/V$ will make the entire Gaussian function lose all $V$ dependence and thus the term will be zero. 
With this, we can now drop the last two terms and recognize that the first term is actually just a sampled operator, namely 

\begin{equation}
    \beta \Pi_{ex} = \left<
    \frac{\partial H[\{V^{-1/2}w_i\}, V^{-1/2d} f]}{\partial V}\right>_{\rm S}.
\end{equation}

At this point, we still need to actually calculate the operator, but we can simply calculate its value for a given simulation state and then sample over its value across a simulation. The operator in question is 
\begin{equation}
\begin{aligned}
    \frac{\partial H[\{V^{-1/2}w_i\}, V^{-1/2d} f]}{\partial V} &= \frac{\partial}{\partial V} \biggl(
    \sum_{i=1}^M \frac{\gamma_i^2}{2 B_i V} 
    \int_\Omega d \rb w_i^{2} (\rb) \\
     &+ \frac{1}{2 E} \int d \rb | \nabla (V^{-1/2d} f(\rb))|^2
     - \sum_j^{P + S + 2} C_j V \log Q_j[\{\mu_i\}, \varphi] \biggr).
\end{aligned}
\end{equation}
Taking a short detour, we will often exploit the general derivative of any extensive property that includes a spatial integral, 
Any extensive variable of the form: 
\begin{equation}
    G = \int d\rb g(\rb)
\end{equation}
will have a dependence on total volume of the system because the spatial integral will vary with the total volume of the system even if the underlying function $g(\rb)$ is invariant to changes in system volume. For any function $G$ that can be defined such that: 
\begin{equation}
\frac{\partial G}{\partial m_p} = 
\frac{\partial V}{\partial m_p} \frac{\partial}{\partial V}
\int d \rb g(\rb)
\end{equation}
We can make the change of variables $\zb = V^{-\frac{1}{d}}\rb$, the area/volume of integration will become invariant under the derivative 
\begin{equation}
\frac{\partial}{\partial V}
\int V d \zb g(V^{\frac{1}{d}}\zb) = 
\int d \zb g(V^{\frac{1}{d}}\zb)
 + \int V d \zb \frac{\partial 
 g(V^{\frac{1}{d}}\zb)}{\partial V}\\
= \frac{1}{V}\int d \rb g(\rb) 
+ \int d\rb \frac{\partial 
 g(\rb)}{\partial V}
\end{equation}

Where the second term will drop out if $g(\rb)$ exhibits no volume dependence. 
Taking each component one at a time and using this special chain/product rule, we have for any i, 
\begin{equation}
    \frac{\partial}{\partial V} 
    \frac{\gamma_i^2}{2 B_i V} 
    \int_\Omega d \rb w_i^{2} (\rb) = 
    - \frac{\gamma_i^2}{2 B_i V^2} 
    \int_\Omega d \rb w_i^{2} (\rb) +  \frac{\gamma_i^2}{2 B_i V^2} 
    \int_\Omega d \rb w_i^{2} (\rb) = 0.
\end{equation}
And for the $\varphi$ term,  
\begin{equation}
    \frac{\partial}{\partial V}\frac{1}{2 E} \int_\Omega d \rb | \nabla V^{\frac{1}{2 d}}\varphi(\rb)|^2 
    =  \frac{1}{2 E V} \int_\Omega d \rb | \nabla \varphi(\rb)|^2 -
    \frac{1}{2 E} \int_\Omega d \rb \frac{1}{V}| \nabla \varphi(\rb)|^2
    = 0 .
\end{equation}
With both of the field terms removed, we can proceed to the partition functions. We already know that 
\begin{equation}
    \frac{\partial}{\partial V} n_j \log Q_j[\{\mu_i\}, \varphi] = 
    n_j Q_j^{-1}\frac{\partial Q_j}{\partial V}
\end{equation}
and can begin with with the simpler solvent case where $Q_j = \frac{1}{V} \int_\Omega d\rb e^{\Gamma * \left(-\psi_j(\rb)\right)}$. In this case we do not need to worry about propagating the volume derivative across the modified diffusion equation, but we still do have to handle the smearing term. We begin by rewriting the partition function with appropriate rescaling (note that we will define $p(\rb) \equiv V^{-\frac{1}{2}} \mu(\rb)$): 

\begin{equation}
    Q_l = \frac{1}{V} \int_\Omega d\rb e^{\Gamma * \left(-V^{-\frac{1}{2}}p_l(\rb) - Z_l V^{-\frac{1}{2d}}f(\rb)\right)/N}
\end{equation}

If we collapse everything in the exponential to $W(V, \rb)$ then we get
\begin{equation}
    \frac{\partial}{\partial V}Q_l = \frac{1}{V} \int_\Omega d\rb \frac{\partial}{\partial V} e^{-W(V,\rb)}
    = \frac{1}{V} \int_\Omega d\rb  e^{-W(V,\rb)} \frac{\partial}{\partial V} (- W(V,\rb)).
\end{equation}

Now we can handle the final term 

\begin{equation}
\begin{aligned}
    \frac{\partial W(V,\rb)}{\partial V} &= \frac{1}{N}\frac{\partial}{\partial V}
    \int_\Omega d \rb' \frac{e^{-\frac{|\rb-\rb'|^2}{2 \alpha^2}}}{(2\pi)^{\frac{d}{2} \alpha^{d}}} V^{-\frac{1}{2}}p_l(\rb) + Z_l V^{-\frac{1}{2d}}f(\rb) \\
    &=\frac{1}{N}
    \int_\Omega d \rb' \frac{1}{V}\left(1 - \frac{|\rb-\rb'|^2}{d \alpha^2}\right)\frac{e^{-\frac{|\rb-\rb'|^2}{2 \alpha^2}}}{(2\pi)^{\frac{d}{2} \alpha^{d}}} V^{-\frac{1}{2}}p_l(\rb) + Z_l V^{-\frac{1}{2d}}f(\rb)\\
    & \quad+ \frac{e^{-\frac{|\rb-\rb'|^2}{2 \alpha^2}}}{(2\pi)^{\frac{d}{2} \alpha^{d}}} \frac{\partial}{\partial V}\left(V^{-\frac{1}{2}}p_l(\rb) - Z_l V^{-\frac{1}{2d}}f(\rb)\right) \\
    &= \frac{1}{VN}
    \int_\Omega d \rb' \left(1 - \frac{|\rb-\rb'|^2}{d \alpha^2}\right)\frac{e^{-\frac{|\rb-\rb'|^2}{2 \alpha^2}}}{(2\pi)^{\frac{d}{2} \alpha^{d}}} V^{-\frac{1}{2}}p_l(\rb) + Z_l V^{-\frac{1}{2d}}f(\rb) \\
    & \quad + \Gamma * (-\frac{1}{2} V^{-\frac{1}{2}}p_l(\rb) + -\frac{1}{2d} Z_l V^{-\frac{1}{2d}}f(\rb)) \\
    &= \frac{1}{NV} (\Gamma_2 - \frac{1}{2}\Gamma) * \psi_l + (\Gamma_2 - \frac{1}{2d}\Gamma) * Z_l \varphi. 
\end{aligned} 
\end{equation} 

This is the same term identified previously \cite{delaney_theory_2017} but with generalizations for dimensionality and in a slightly different basis. Where we have defined $\Gamma_2(\rb) = (1 - \frac{|\rb|^2}{d \alpha^2}) \Gamma(\rb)$. Conveniently, $\Gamma_2(\textbf{k}) = \alpha^2 |\textbf{k}|^2 \Gamma(\textbf{k})$ in Fourier space, regardless of dimensionality. This is also the correct solution for the salts with the usual salt condition of $\psi(\rb) = 0$. 

Putting everything together, we get for the solvent that 
\begin{align}
    \frac{\partial}{\partial V} C_l V \log Q_l[\{\mu_i\}, \varphi] & = 
   -C_l V Q_l^{-1} \frac{1}{NV} \frac{1}{V} \int_\Omega d\rb  e^{-W(V,\rb)}  \left((\Gamma_2 - \frac{1}{2}\Gamma) * \psi_l(\rb) + (\Gamma_2 - \frac{1}{2d}\Gamma) * Z_l \varphi(\rb)\right) \nonumber \\
   & = -\frac{1}{V}\int_\Omega d \rb \rho_l(\rb) \left((\Gamma_2 - \frac{1}{2}\Gamma) * \psi_l(\rb) + (\Gamma_2 - \frac{1}{2d}\Gamma) * Z_l \varphi(\rb)\right). 
\end{align}
Conveniently, this is a product of the convolved field and the density, which are all commonly used operators, so this extra operator is fairly cheap to compute. 

\subsubsection{Modified Diffusion Derivative}

While the polymer case is quite similar, we do have to calculate how the modified diffusion equation changes with changes in volume, which is somewhat involved. Again, this derivation is essentially a generalization of previous results, done in full for clarity. We begin by writing down our modified diffusion equation with the correctly rescaled variables, namely

\begin{equation}
    \frac{\partial}{\partial s} q(V^{1/d} \zb, s, W(s, \rb)) 
    = V^{-2 / d} \nabla_{\zb}^2 q(V^{1/d} \zb, s, W(s, \rb)) 
    - W(s, V^{1/d} \zb) q(V^{1/d} \zb, s, W(s, \rb)) . 
\end{equation}

We then take the derivative of both sides with respect to $V$, 
\begin{equation}
    \frac{\partial^2 q}{\partial s \partial V} 
    = \nabla^2 \frac{\partial q}{\partial V} - \frac{2}{d V} \nabla^2 q
    - W(s, \rb) \frac{\partial q}{\partial V} - \frac{\partial W(s, \rb)}{\partial V} q
\end{equation}

and rearrange to get a nonhonogeneous equation in $\frac{\partial q}{\partial V}$ 
\begin{equation}
    \left(\frac{\partial}{\partial s } - \nabla^2 + W(s, \rb)\right) \frac{\partial q}{\partial V}
    = - \frac{2}{d V} \nabla^2 q
    - \frac{\partial W(s, \rb)}{\partial V} q
\end{equation}

with the homogeneous condition that $\frac{\partial q(\rb, 0)}{\partial V} = 0$.  

This can be solved with the associated Green's function 

\begin{equation}
    \frac{\partial g}{\partial s} + \mathcal{L}_{\rb} g = 0;
    \mathcal{L}_{\rb} \equiv - \nabla_{\rb}^2 + W(s, \rb);
    g(\rb, \rb', 0) = \delta(\rb - \rb').
\end{equation}

where the periodic boundary conditions can be formally expressed as 

\begin{equation}
    g(\rb, \rb', s, W(s, \rb)) 
    = e^{-\mathcal{L}_{\rb} s}\delta(\rb - \rb').
\end{equation}

Our original propagator can be related to the Green's function as

\begin{equation}
    q(\rb, s, W(s, \rb)) = \int_\Omega d \rb' g(\rb, \rb', s; W(s, \rb)).
\end{equation}

Because $q^{\dagger}$ has the inverse causality of $q$ but otherwise is identical, there is no problem in identifying 
\begin{equation}
    q^{\dagger}(\rb, s^* - s, W(s, \rb)) = \int_\Omega d \rb' g(\rb', \rb, s; W(s, \rb)). 
\end{equation}

From here we can construct solution to the nonhomogenous equation using a forcing function, $F(\rb, s)$

\begin{equation}
    \frac{\partial f(\rb, s)}{\partial s} + \mathcal{L}_{\rb} = F(\rb, s),
\end{equation}
\begin{equation}
    e^{-\mathcal{L}_{\rb} s} \frac{\partial}{\partial s} 
    (e^{\mathcal{L}_{\rb} s} f(\rb, 0)) = \int_\Omega d \rb' F(\rb, s) \delta(\rb - \rb'),
\end{equation}
\begin{equation}
    f(\rb, s) - e^{-\mathcal{L}_{\rb} s} f(\rb, 0)
    = \int_0^{s} d s' \int_\Omega d \rb' F(\rb', s')
    e^{-\mathcal{L}_{\rb} (s - s')} \delta(\rb - \rb').
\end{equation}

We can simplify with our initial condition and write our Green's function explicitly as 
\begin{equation}
    f(\rb, s) 
    = \int_0^{s} d s' \int_\Omega d \rb' F(\rb', s')
    g(\rb, \rb', s - s').
\end{equation}

Next we replace the Green's function with our inverse propagator to get

\begin{align}
    \int_\Omega d \rb f(\rb, s) 
    & = \int_0^{s} d s' \int_\Omega d \rb' F(\rb', s')
    \int_\Omega d \rb g(\rb, \rb', s - s') \nonumber \\
    & = \int_0^{s} d s' \int_\Omega d \rb F(\rb, s')
    q^{\dagger}(\rb, s, W(s, \rb)). 
\end{align}

By using the previously established definition for the forcing term, we can now write
\begin{equation}
\begin{aligned}
    \int_\Omega d \rb \frac{\partial q(\rb, s)}{\partial V} 
    & = \int_0^{s} d s' \int_\Omega d \rb 
    q^{\dagger}(\rb, s', W(s, \rb)) \\
    & \quad * \left(\frac{2}{d V} \nabla^2 + \frac{\partial W(s, \rb)}{\partial V})\right) q(\rb, s', W(s, \rb))
\end{aligned}
\end{equation}
which conveniently is closely related to the density operator, 
\begin{equation}
    \frac{\partial Q[W]}{\partial V} = -\frac{Q}{V} \int_\Omega d \rb \int_0^{\frac{N_j}{N}} d s \frac{2}{V d}\rho_{\nabla j}(\rb,s) + \rho_j(s) \frac{\partial W(\rb, s)}{\partial V}
\end{equation}
where 
\begin{equation}
    \rho_{\nabla j}(\rb,s) \equiv Q^{-1} q(\rb, s) \nabla^2 q^{\dagger} (\rb, s)
\end{equation}
and $\rho$ uses the conventional definition of 

\begin{equation}
    \rho_{j}(\rb,s) \equiv Q^{-1} q(\rb, s) q^{\dagger} (\rb, s).
\end{equation}

At this point we have everything to finish the derivative, we can use the same $\frac{\partial W}{\partial V}$ as in the solvent case to get: 

\begin{equation}
\begin{aligned}
    \frac{\partial}{\partial V} C_j V \log Q_j[\{\mu_i\}, \varphi] &=
   - \frac{1}{V}\int_\Omega d \rb \int_0^{\frac{N_j}{N}} d s \biggl[ 
   \frac{2}{d}\rho_{\nabla j}(\rb,s) \\
   &+
   \rho_j(\rb, s) \left((\Gamma_2 - \frac{1}{2}\Gamma) * \psi_l(\rb, s) + (\Gamma_2 - \frac{1}{2d}\Gamma) * Z_l \varphi(\rb, s) \right)\biggr].
\end{aligned}
\end{equation}
\subsubsection{Ideal Gas and Final result}

The only thing that we have left to calculate is the ideal gas derivative, which is straightforward. We just have 
\begin{equation}
    \frac{\partial}{\partial V} n_j \log V = \frac{n_j}{V} = C_j.
\end{equation}
With this we can write down our final operator, namely 
\begin{equation}
\begin{aligned}
    \beta \Pi &= \sum_j^{P + S + 2} C_j    -\frac{\textbf{c}^T \chi \textbf{c}}{2} \\
    & \quad + \sum_j^P \frac{1}{V}\int_\Omega d \rb \int_0^{\frac{N_j}{N}} d s \biggl[
   \frac{2}{d}\rho_{\nabla j}(\rb,s) \\
   & \quad + \rho_j(\rb, s) 
   \left((\Gamma_2 - \frac{1}{2}\Gamma) * \psi_j(\rb, s) + (\Gamma_2 - \frac{1}{2d}\Gamma) * Z_j \varphi(\rb, s)\right)\biggr]\\
   & \quad + \sum_j^{S+2} \frac{1}{NV}\int_\Omega d \rb \rho_j(\rb) \left((\Gamma_2 - \frac{1}{2}\Gamma) * \psi_j(\rb) + (\Gamma_2 - \frac{1}{2d}\Gamma) * Z_j \varphi(\rb)\right) 
\end{aligned}
\end{equation}

\subsection{Gibbs Ensemble}

With both the osmotic pressure and chemical potentials in hand, we can now build our Gibbs ensemble. In this work we only used the Gibbs ensemble to validate our model by recreating previous results, so we will restrict ourselves to models that only consider implicit solvent and are free of explicit salt. This condition simplifies all the previous equations by neglecting any contribution not related to polymer. This also simplifies the Gibbs ensemble because there is only one charge neutral move that is possible. Because each simulation must be charge neutral, every move must also be charge neutral. With mixtures of 3 or more charged species it is not always obvious how to correctly determine the best charge neutral move, but for a 2 component mixture there is only one option. With these preliminaries noted, we can proceed to write down our Gibbs ensemble dynamics. 

We begin by instantiating to simulations which can be at any arbitrary state, but ideally will be selected to be near the two pure phases we are trying to demonstrate coexistence for. These simulations will each have a volume ($V_{I}$ and $V_{II}$) and concentration ($C_{Ij}$ and $C_{IIj}$ for each species). We can also determine the amount of particles in each simulation, for convenience denoted $m_{Ij}$ and $m_{IIj}$ for which $m_{Ij} = C_{Ij}V_I$ and so on. The total mass and volume are conserved so that $V_I + V_{II} = V_T$, and $m_{Ij} + m_{IIj} = m_{Tj}$ for all species. The joint free energy of the system can be written as $F_I({n_{Ij}}, V_I, T) + F_{II}({n_{IIj}}, V_{II}, T) = F({n_{Tj}}, V_T, T)$. The system will be in equilibrium if 

\begin{equation}
     \frac{\partial F}{\partial n_{Ij}} = (\mu_I - \mu_{II}) = 0 
\end{equation}
for all $j$ and 
\begin{equation}
     -\frac{\partial F}{\partial V_I} = (\Pi_I - \Pi_{II}) = 0 
\end{equation}
where $\mu$ and $\Pi$ are the chemical potential and osmotic pressure calculated previously. This equilibrium condition can be used to write simple equations of motion

\begin{equation}
    \frac{\partial m_{Ij}}{\partial t} = \mu_{IIj} - \mu_{Ij}
\end{equation}
\begin{equation}
    \frac{\partial V_{I}}{\partial t} = \Pi_{I} - \Pi_{IIj}
\end{equation}

with first order update scheme 

\begin{equation}
    m_{Ij}(t + \Delta t) = m_{Ij}(t) + \Delta t_j \left(\mu_{IIj} - \mu_{Ij}\right)
\end{equation}
\begin{equation}
     V_{I}(t + \Delta t) = V_{I}(t) + \Delta t_V \left(\Pi_{I} - \Pi_{II}\right).
\end{equation}

Our update scheme operates on the mass and volumes, and the concentrations can be trivially recovered from their values for each cell. In principle the update step for each species can be chosen separately, but for the simple two-component case we will replace the two $m_{Ij}$'s with a single mass and  chemical potential for the charge neutral pair $m_I = Z_1 m_{I1} - Z_2 m_{I2}$ and  $\mu_I = Z_1 \mu_{I1} - Z_2 \mu_{I2}$. For the polyampholyte, the polymer is charge neutral and there is only one species. 

The update scheme is allowed to proceed until a stationary point is found, at which point phase coexistence is proven. In principle, higher order update schemes could be used to achieve faster convergence. Because the Gibbs ensemble is not a significant focus of this work and is only used to demonstrate that our model is consistent with known results, we have not examined these potential improvements. 

\subsection{Comparison to Previous Work}

With all the mathematical preliminaries handled, we now present the direct comparisons between our model and previous simulation methods of coacervates \cite{delaney_theory_2017}. 
We present two results here using the same conditions as Delaney's previous work which show good agreement with our current methods. 

\begin{figure}[h]
    \centering
    \includegraphics{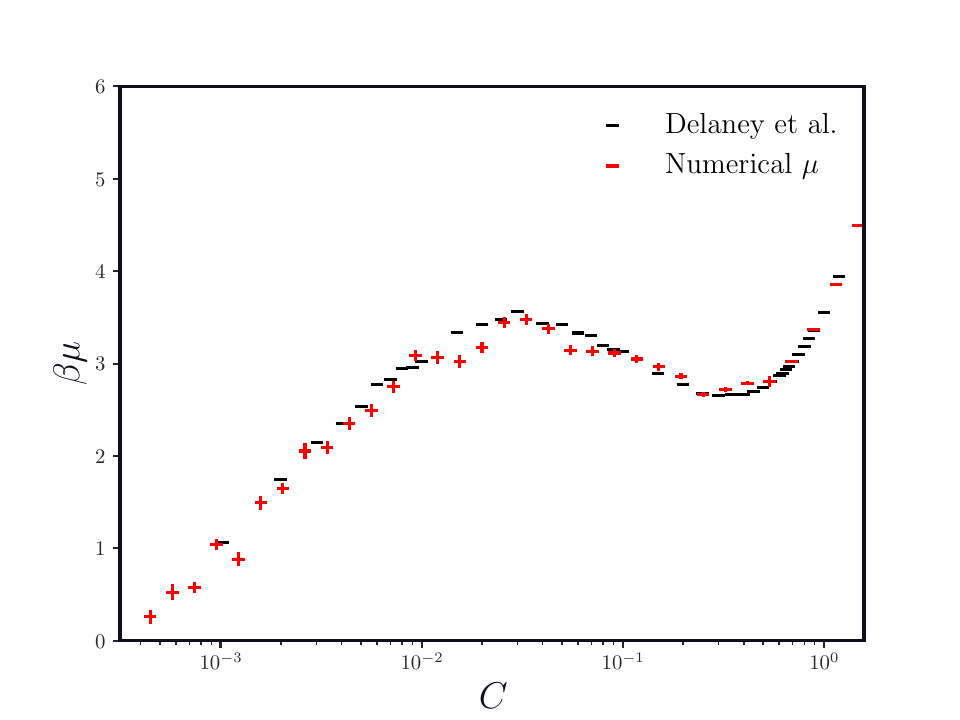}
    \caption{Chemical potential versus concentration for use in phase equilibrium conditions. Simulation is a diblock polyampholyte with $B=1.0$, $E=400$, $a=0.2$. Comparison is against figure 5b in \cite{delaney_theory_2017}.
    \label{fig:replication_mu}}
\end{figure}

First we were able to make a close match on the chemical potential of a polyampholyte simulation at different polyampholyte concentrations. 
While there is some variation between the two plots, this can be attributed to differences in dynamic parameters and random number generation. 
The estimates of variance likely underestimate the true variance due to non-ergodic sampling that is difficult to capture with standard methods used to identify the variance compared to the true underlying distribution. 

\begin{figure}
    \centering
    \includegraphics{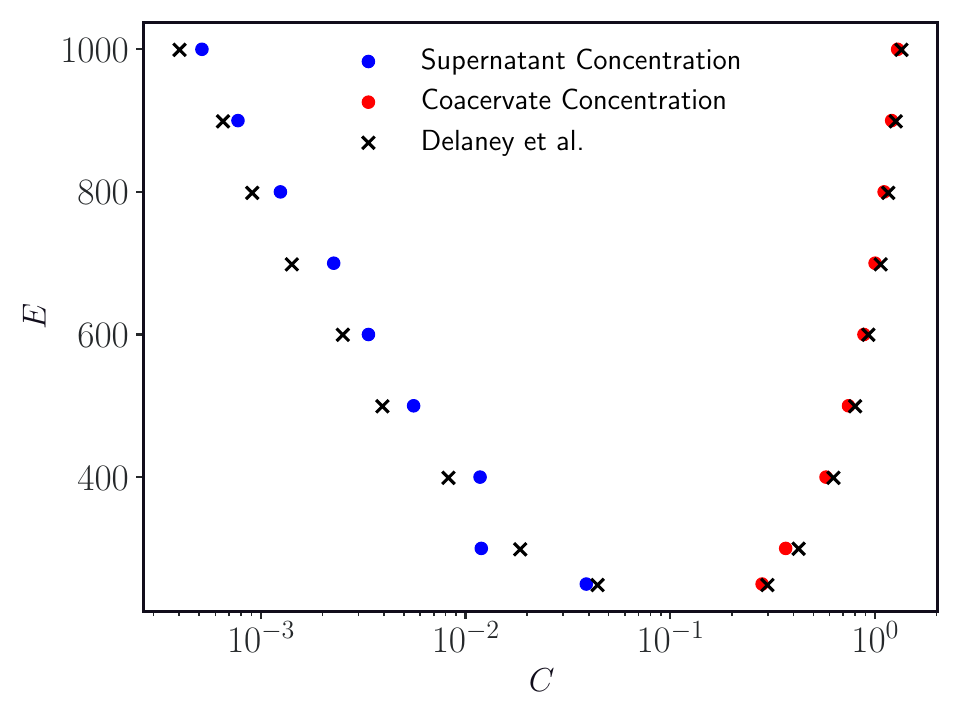}
    \caption{Plots of phase coexistence for a symmetric diblock polyampholyte at various values of E. Comparison is to figure 6 of \cite{delaney_theory_2017} using $a=0.2$ and $B=1$ \cite{shirts_statistically_2008} \cite{chodera_simple_2016}.
    \label{fig:replication_coexistence}}
\end{figure}

We also made comparisons against the final phase concentration predictions made by running Gibbs ensemble simulations as shown in Fig. \ref{fig:replication_coexistence}. 
Here we show close agreement for the density of the coacervate phase and reasonably good agreement on the supernatant phase. 
Measuring the exact density of the supernatant phase is particularly difficult due to the small concentrations in question. 
These small values exacerbate the variance of the noise relative to the signal and amplify issues with non-ergodic sampling. 
Still we believe that we have shown sufficiently close agreement overall to indicate that our simulation method correctly recapitulates previously reported results. 
Without access to the underlying code used to generate these previous results, there is no practical way to discern small differences in performance and we have shown that we achieve the same results as previous work. 

\section{Notation Glossary}
\subsection{System-wide Constants}
\begin{tabular}{c|c}
     $M$ & number of monomer species \\
     $P$ & number of polymer types \\ 
     $S$ & number of solvent types \\ 
     $N$ & reference polymer length \\
     $b$ & reference bond stretching parameter \\
     $R_{\rm g}$ & reference radius of gyration \\
     $d$ & system dimensionality \\
     $V$ & rescaled volume \\
     $\Delta V$ & rescaled volume of unit cell \\
     $a$ & smearing constant \\
     $E$ & rescaled Bjerrum length \\
\end{tabular}

\subsection{Spatial Variables and functions}
\begin{tabular}{c|c}
     $\rb$ & rescaled position vector \\
     $s$ & scaled polymer space curve coordinate \\
     $\omega_i(\rb)$  & chemical potential field in the diagonalized basis \\
     $\mu_i(\rb)$  & $\omega_i$ after Wick rotation \\
     $\varphi(\rb)$ & electrical potential field \\

     $\psi_i(\rb)$ & Field experienced by monomer type i \\
     $\rho_{i}(\rb)$ & Density vector for monomer type $i$ \\
     $q_{p}(\rb,s)$ & forward chain propagator for polymer type $p$ \\
     $q_{p}^{\dagger}(\rb, s)$ & reverse chain propagator for polymer type $p$ \\

     $\Gamma$ & Gaussian smearing kernel \\

\end{tabular}
\subsection{Indexed Parameters}
\begin{tabular}{c|c}
     $n_{p}$ & number of polymers of species $p$ \\
     $N_{p}$ & length of polymers of species $p$ \\
     $C_j$ & reduced concentration of species $j$\\
     $F_{ij}$ & Flory-Huggins interaction between monomers of types $i$ and $j$ \\ 
     $B_i$ & Diagonalized Flory-Huggins interaction for the $i$ diagonalized field \\
     $m_{p}(s)$ & monomer identity of polymer type $p$ at $s$ \\
     $m_{s}$ & monomer identity of solvent type $s$ \\
     $Z_{m}$ & charge per monomer of type $m$ \\
     $\gamma_i$ & $\omega_i$ Wick rotation variable \\
     $Q_j$ &single unit partition function for species type $j$ \\
\end{tabular}

\subsection{Dynamics Variables}

\begin{tabular}{c|c}
     $\lambda_{j\mu}$ or $\lambda_{\varphi}$ & step size for corresponding dynamics\\
     $\eta_{j\mu}$ or $\eta_{\varphi}$ & random noise injected into corresponding dynamics\\
     $\beta_{j\mu}$ or $\beta_{\varphi}$ & fictitious temperature for corresponding dynamics\\
     $t$ & fictitious time \\
     $\Delta t$ & fictitious time step \\  
     $c(\kb)$ & analytical approximation of linear response of force to fields\\

\end{tabular}
\subsection{Analytic Constant terms and Partition Functions}

\begin{tabular}{c|c}
     $\bb$ & matrix of eigenvalues to transform species basis to diagonalized basis\\
     $\kappa_{ij}$ & fraction of $i$-th species composed of the $j$-th monomer type \\
     $\cb$ & average concentration vector \\
     $\mathcal{Z}$ & total partition function \\
     $\mathcal{Z}_{\rm ideal}$ & ideal gas contribution to partition function \\
     $\mathcal{Z}_{\rm C}$ & charge contribution to partition function \\
     $\mathcal{Z}_i$ & diagonal field $i$ contribution to the parition function \\

\end{tabular}
\subsection{Gibbs ensemble variable}

\begin{tabular}{c|c}
     $\mu_i$ & chemical potential for species $i$   \\     
     $m_i$ & total mass of species $i$ \\

     $\Pi$ & osmotic pressure \\
     $V$ & volume \\
     $F$ & free energy \\
     $\Delta t$ & fictitious time step
\end{tabular}

All species are in either simulation $I$ or $II$. The corresponding net values are denoted with a $\Delta$.

\end{document}

%% file: preamble.tex
\usepackage[x12names, rgb, html]{xcolor}

\definecolor{red}{HTML}{f54b1a}
\definecolor{pink}{HTML}{d19eb1}
\definecolor{orange}{HTML}{d3772e}
\definecolor{yellow}{HTML}{ebe85d}
\definecolor{green}{HTML}{0f6852}
\definecolor{lightblue}{HTML}{01abe9}
\definecolor{darkblue}{HTML}{1b346c}
\definecolor{tan}{HTML}{e5c39e}
\definecolor{darktan}{HTML}{af9e73}
\definecolor{grey}{HTML}{c3ced0}
\definecolor{darkgrey}{HTML}{9dadc4}
\definecolor{black}{HTML}{110d1b}
\definecolor{white}{HTML}{f1f8f1}

\usepackage{hyperref}
\hypersetup{
  colorlinks = true,
  linkcolor = red,
  citecolor = darkblue,
  urlcolor = lightblue
}

\usepackage[cmintegrals,cmbraces]{newtxmath}

\usepackage{graphicx}
\usepackage{adjustbox}

\usepackage[margin=1in]{geometry}

\usepackage{algorithm}
\usepackage{algpseudocode}
\algrenewcommand{\algorithmiccomment}[1]{$\vartriangleright$ #1}
\algrenewcommand{\algorithmicreturn}{\textbf{Return: }}
\algnewcommand\algorithmicinput{\textbf{Input: }}
\algnewcommand\Input{\State \algorithmicinput}

\def\rhob{\boldsymbol{\rho}}

\def\bb{\boldsymbol{b}}
\def\cb{\boldsymbol{c}}

\def\kb{\boldsymbol{k}}
\def\rb{\boldsymbol{r}}

\def\xb{\boldsymbol{x}}

\def\zb{\boldsymbol{z}}

\def\d{\displaystyle}

\def\RR{\mathbb{R}}

\def\<{\langle} \def\>{\rangle}